\documentclass[twocolumn]{aastex631}

\usepackage{rotating}
\usepackage{multirow}

\begin{document}

\title{A Systematic Search for X-ray Eclipse Events in Active Galactic Nuclei Observed by {\it Swift}}

\author[0009-0001-7373-5388]{Tianying Lian}
\affiliation{National Astronomical Observatories, Chinese Academy of Sciences, 20A Datun Road, Beijing 100101, People’s Republic of China}
\affiliation{School of Astronomy and Space Science, University of Chinese Academy of Sciences, 19A Yuquan Road, Beijing 100049, People’s Republic of China}

\author[0000-0002-2006-1615]{Chichuan Jin}
\affiliation{National Astronomical Observatories, Chinese Academy of Sciences, 20A Datun Road, Beijing 100101, People’s Republic of China}
\affiliation{School of Astronomy and Space Science, University of Chinese Academy of Sciences, 19A Yuquan Road, Beijing 100049, People’s Republic of China}
\affiliation{Institute for Frontier in Astronomy and Astrophysics, Beijing Normal University, Beijing 102206, China}

\author{Weimin Yuan}
\affiliation{National Astronomical Observatories, Chinese Academy of Sciences, 20A Datun Road, Beijing 100101, People’s Republic of China}
\affiliation{School of Astronomy and Space Science, University of Chinese Academy of Sciences, 19A Yuquan Road, Beijing 100049, People’s Republic of China}



\begin{abstract}

The nuclear regions of active galactic nuclei (AGNs) likely host clumpy clouds that occasionally obscure the central X-ray source, causing eclipse events. These events offer a unique opportunity to study the properties and origins of such clouds. However, these transient events are rarely reported due to the need for extensive, long-term X-ray monitoring for years. Here, we conduct a systematic search for eclipse events in 40 AGNs well-monitored by the \textit{Swift} X-ray Telescope (XRT) over the past 20 years, comprising a total of $\sim$11,000 observations. Our selection criteria rely on significant variations in X-ray flux and spectral shape. We identify 3 high-confidence events in 3 AGNs and 8 candidates in 6 AGNs, all in type I AGNs. The observed clouds have column densities of $N_{\rm H}~\sim$ (0.2 $-$ 31.2) $\times$ 10$^{22}$ cm$^{-2}$ and ionization degrees of log $\xi$ $\sim$ (-1.3 $-$ 2.2). For the 5 events with well-constrained duration, their distances from the central black hole range from (2.4 $-$ 179) $\times$ 10$^{4}$ $R_{\rm g}$, with 2 clouds near the dust sublimation zone, 2 farther out. Interestingly, we find tentative correlations between the cloud properties (i.e. ionization state and column density) and the black hole mass and mass accretion rate, implying their strong connection to the accretion process, potentially via outflows. Our study also demonstrates the potential of the new X-ray all-sky monitor {\it Einstein Probe} in providing more detection and physical constraints for such events.

\end{abstract}

\keywords{X-ray active galactic nuclei (2035)}


\section{Introduction} \label{sec:intro}

\defcitealias{2014MNRAS.439.1403M}{M14}

Powered by the accretion of gas onto central supermassive black holes (SMBHs), active galactic nuclei (AGNs) emit radiation across almost the entire electromagnetic spectrum with variations at various timescales. Previously, both observational and theoretical studies suggested that clumpy clouds may exist around AGNs. These clouds are thought to originate from outflows \citep[e.g.,][]{2000ApJ...543..686P,2001ApJ...561..684K}, broad line region (BLR) \citep[e.g.,][]{1997MNRAS.288.1015A,2011MNRAS.410.1027R}, or dusty torus \citep[e.g.,][]{2002ApJ...570L...9N,2008A&A...482...67S}. These clouds can alter the line-of-sight column density and obscure the radiation from the central engine, causing variations in both flux and spectral shape across optical/UV and X-ray bands \citep[e.g.,][]{2003MNRAS.342L..41L,2009MNRAS.393L...1R}. Previous studies have indicated that optical/UV obscuring clouds are typically dusty, whereas X-ray clouds can be either dusty or dust-free \citep[e.g.,][]{2001A&A...365...28M,2014MNRAS.439.1403M,2020MNRAS.493..930J}. It was also suggested that these two types of obscuration can be associated with the same obscuring material \citep[e.g.,][]{2014Sci...345...64K,2015ARA&A..53..115K,2021ApJ...922..151K}. To investigate these clouds and understand their physical properties, one effective approach is to identify their distinct pattern of flux variation, observed as an eclipse event \citep[e.g.,][]{2003MNRAS.342L..41L,2009MNRAS.393L...1R,2010A&A...517A..47M}. However, due to the transient nature of these eclipse events, their detection requires high-cadence, long-term monitoring of AGNs.

Thanks to the long-term monitoring data accumulated by various telescopes in the past decades, more and more eclipse events have been identified in AGNs, especially in X-rays. Some eclipse events were captured and further studied with high-quality spectra \citep[e.g.][]{2016MNRAS.456L..94M,2021ApJ...908L..33G,2023MNRAS.525.1941K}. Other events were identified by the archival observations with lower data quality. For example, \citet{2014MNRAS.439.1403M} (hereafter:\citetalias{2014MNRAS.439.1403M}) conducted a sample search based on the long-term data observed by \textit{Rossi X-ray Timing Explorer} (\textit{RXTE}), and identified 12 secured eclipse events in type I and II AGNs. Their results suggested that the eclipse gases found by \textit{RXTE} are likely dusty. \citet{2014MNRAS.442.2116T} used \textit{XMM-Newton} and \textit{Suzaku} observations of 42 brightest sources to search for occultations. They reported that 15 sources exhibited highly probable eclipse events, likely caused by BLR clouds. Another sample study, consisting of 6 AGNs observed by \textit{XMM-Newton}, also revealed several eclipses originating from BLR \citep{2022MNRAS.514.1535A}. These studies primarily rely on the variations in the hardness ratio light curve, while others identify eclipses by searching for the unusual trend in the hardness ratio (HR) $-$ count rate (CR) diagram \citep[e.g.][]{2023MNRAS.525.1941K}.

Additionally, previous observations indicated that the timescales of eclipse events range from hours to years, partially or fully covering the intrinsic emission with neutral or modestly ionized gases during the period. The eclipse clouds can be located in the BLR, inside or outside the dust sublimation zone (DSZ) \citep[e.g. ][]{2010A&A...517A..47M,2014MNRAS.439.1403M,2023MNRAS.525.1941K}. Some AGNs also exhibit more than one eclipse event during their historical observation intervals, providing further clues for understanding the occurrence rate of such events\citep[e.g. NGC 3227:  ][]{2003MNRAS.342L..41L,2015A&A...584A..82B,2018MNRAS.481.2470T,2022A&A...665A..72M}.

These observational discoveries promote the establishment of new theoretical models. The initial formalism for the clumpy torus models was progressively developed \citep[e.g.][]{2002ApJ...570L...9N,2004ASPC..320..242E,2008ApJ...685..147N,2008ApJ...685..160N}. This CLUMPY model incorporates dusty clouds distributed along the equatorial plane to construct the torus, which effectively explains the infrared observations of AGNs \citep[e.g.][]{2004ApJ...616..123T,2006ApJ...640..612M,2014A&A...567A.125G}. Subsequently, many studies adopted a geometry similar to that in the CLUMPY model to explain the X-ray observations. \citetalias{2014MNRAS.439.1403M} first extended this model to the X-ray band, independently constraining the model's parameter space with their observational results of the eclipse clouds. An X-ray clumpy torus model named XCLUMPY was developed to explain the X-ray spectrum of Circinus galaxy and constrain the torus covering factors \citep{2019ApJ...877...95T}, which has been applied to many studies \citep[e.g.][]{2019ApJ...884L..10M,2021ApJS..257...61Y}. To better interpret the variable clumpy structures that cause eclipse events and predict their occurrence rate, \citet{2019A&A...629A..16B} developed the unified clumpy model UXCLUMPY. This model is based on their Monte Carlo X-ray radiative transfer code (XARS) and was applied to analyze X-ray spectra from \textit{NuSTAR} data, which demonstrates a self-consistent combination with the CLUMPY model.

To provide sufficient observational constraints and tests for the existing models, aiding our understanding of the properties of clouds around these AGNs, a larger sample of eclipse events is needed. However, the current sample of X-ray eclipse events remains limited. {\it Swift} \citep{2004ApJ...611.1005G} has monitored many AGNs in X-rays for about 20 years with high sensitivity, thereby providing us a good opportunity to search for new X-ray eclipse events.

In this paper, we conduct a systematic study of an AGN sample with long-term monitoring observations by {\it Swift} X-ray Telescope (XRT), as described in Section \ref{sec-2}. Then in Section \ref{sec-3} we describe our strategy for searching for eclipse events, specifically designed to accommodate the characteristics of \textit{Swift} data. The newly discovered eclipse events are detailed in Section \ref{sec:results}. Further discussions such as the physical properties of such events, and comparisons with previous works are given in Section \ref{sec:discussion}. The final Section \ref{sec:summary} summarizes all the results.

\section{The AGN Sample and Data}
\label{sec-2}
\subsection{Sample Selection}
\label{sec:sample}
Searching for eclipse events requires identifying characteristic variation patterns from the normal long-term variation of AGN. \textit{Swift} has conducted long-term monitoring observations for a sample AGN in the past decades, providing a rich database for the searching of eclipse events.

In this paper, we first use two previously reported AGN samples to construct our preliminary large sample, including 78 Seyfert galaxies observed by the \textit{XMM-Newton}
\citep{2001A&A...365L...1J} from \citet{2022ApJ...936..105H} and 153 AGN from \textit{RXTE} AGN Timing \& Spectral Database \citep{2013ApJ...772..114R}. This candidate AGN list is mainly composed of bright and nearby AGNs, and some of them are radio-loud AGNs. 

To better define the normal variation of AGN to select the eclipse event, we only consider sources that have been observed by \textit{Swift}/XRT for more than 50 times and adopt the XRT observations of these sources where the detection significance was large than 3$\sigma$. Then we calculate the hardness ratio and perform further exploration of eclipse events. In addition, we exclude radio-loud AGNs whose X-ray emission is dominated by jets.

Our final sample consists of 40 AGNs. The XRT observations of each source with its total exposure time and redshift are listed in Appendix \ref{app:A}. The black hole mass ($M_{\rm BH}$) of these sources compiled from the literature are listed in Table \ref{tab:appA}\footnote{We prioritize black hole mass estimates in the following order: first from GRAVITY, followed by reverberation mapping, then from empirical relations between black hole mass and either large-scale galaxy properties or AGNs' emission line properties, or virial masses measured from the broad-line region.}. The redshifts of the sample span from 0.001 to 0.237, with an average value of 0.034. Noteworthy, our sample also includes some well-known changing-look AGNs such as 1ES 1927+654 \citep{2019ApJ...883...94T} and ESO 323-G77 \citep{2014MNRAS.437.1776M}. For this AGN sample, we go through the literature and find that some obscured events have been reported before, which showed the duration from hours to several months. Some AGNs have more than one obscuration event. For example, NGC 1365 was identified several eclipse events in 2004, 2006, 2007, and between 2012 and 2013 \citep{2007ApJ...659L.111R,2009ApJ...696..160R,2010A&A...517A..47M,2014ApJ...788...76W}. These known eclipse events are also summarized in Appendix \ref{app:A}.

\subsection{Data Reduction}
\label{sec:observation}
We perform local data reduction for the \textit{Swift} XRT and Ultra-violet Optical Telescope (UVOT) data by constructing a pipeline. We download the \textit{Swift} data from NASA's High Energy Astrophysics Science Archive Research Center (HEASARC). All the \textit{Swift} data were processed with the \texttt{HEADAS/Swift} software version \texttt{5.8}, \texttt{FTOOLS} version \texttt{6.32.1} and \texttt{CALDB} version \texttt{20220803}. The details of processing the XRT and UVOT data and calculating the hardness ratio are described below. 

\subsubsection{XRT Data}
\label{sec:XRTdata}
The data from each observation is firstly processed by \texttt{xrtpipeline}. Observations in both the photon counting (PC) and windowed timing (WT) modes are adopted. Events with grade 0–12 and 0–2 are used for the PC and WT modes, respectively. For observations conducted in multiple XRT modes, we adopt the one with the longest exposure time. Then our pipeline calculates the significance of source detection. Observations below 3$\sigma$ detection are excluded. The upper limit of source detection is not used in this study because the data cannot constrain the hardness ratio of the source.

As the standard procedure for extracting both PC and WT modes data, circles and annulus are used to define the source and background regions with the center of the background annuls matching the center of the circular region. If in the PC mode the source count rate is $>$ 0.5 counts s$^{-1}$, an annular region for source extraction is used to avoid the pile-up effect. The choice of the inner radius for the source extraction region is based on a comparison between the observed point-spread function (PSF) and the theoretical PSF to determine the radius within which pile-up may occur. This radius is then set as the inner radius of the source extraction region. The outer radius is typically selected within the range of 60 $-$ 120 arcsec, depending on the brightness of the source. For brighter sources, a larger outer radius is chosen to include more source photons.

The light curve and spectrum are extracted with \texttt{XSELECT} version \texttt{2.4k}. The Ancillary Response Files(ARFs) are created by using \texttt{xrtmkarf} after producing the exposure map with \texttt{xrtexpomap}. The ARFs contain the effective area of XRT in the energy grid which is read from the Response Matrix files, and the value of effective area includes the vignetting correction and the PSF correction. Then we calculate the count rate for each energy grid, normalize it by the corresponding effective area, and subsequently aggregate them to obtain the total flux in units of counts s$^{-1}$ cm$^{-2}$ for a single observation. With this method, our pipeline generates light curves for every source in the sample, with the initial time bin set as one observation per bin. Additionally, we use \texttt{addspec} to stack spectra obtained during selected time intervals and \texttt{grppha} to prepare spectra for fitting in \texttt{XSPEC}.

\subsubsection{UVOT Data}
\label{sec:UVOTdata}
We use the \texttt{uvotsource} tool to measure the flux densities in six filters (i.e. UVW2, UVM2, UVW1, U, B and V) for individual exposures. The flux densities are extracted using a circular source region with a radius of 5 arcsec, while the background is determined from a nearby region with a radius of 25 arcsec. We set the parameter \texttt{ssstype = low} within the \texttt{uvotsource} tool to identify and exclude some data where the source happened to locate on a low-sensitivity area of the UVOT detector. Additionally, exposures that fail the satellite pointing stability test are also excluded from further analysis. Each data point on the UVOT light curve is computed as an exposure-weighted average, with statistical errors propagated accordingly.

\section{Searching for Eclipse Events}
\label{sec-3}

\subsection{Searching Strategy and Criteria}
\label{sec:3.1}

Our strategy for searching and identifying eclipse events is primarily based on the method described in \citetalias{2014MNRAS.439.1403M}, which was applied to the searching of eclipse events in the {\it RXTE} data. We make various modifications to the method so that it can be more suitable for the XRT data. Our final method involves a comprehensive evaluation of each source’s flux light curve, hardness ratio curve, and spectral parameter curves. The hardness ratio is defined as (F$_{\rm 2-10 keV}$ $-$ F$_{\rm 0.3-2 keV}$)$/$(F$_{\rm 2-10 keV}$ + F$_{\rm 0.3-2 keV}$). It is derived by applying the program named Bayesian estimation of hardness ratios \citep[BEHR,][]{2006ApJ...652..610P}.

For each source, we apply the following four selection criteria to identify an eclipse event:

(1) Criterion 1: The hardness ratio observed in at least two consecutive observations exceeds the average hardness ratio across all observations, with the deviation exceeding 2$\sigma$.

(2) Criterion 2: The flux observed during this high hardness ratio period should be lower than the average value over the entire observations.

These two criteria reflect the typical characteristics of an eclipse event, where the X-ray flux drops abruptly over a specific period, accompanied by spectral hardening due to additional absorption components. We also note that Criterion 1 is relatively stringent, as we cannot know in advance which period an eclipse occurred. Therefore, the standard deviation of the hardness ratio is calculated using the entire light curve, including all the potential eclipse events, which can increase the standard deviation. Therefore, events selected by Criterion 1 are expected to be the most prominent ones. 

A period satisfying both two criteria is defined as a potential eclipse phase. Additionally, we select a nearby period outside the eclipse phase where the hardness ratio remains stable and close to its average value, defining it as the non-eclipse phase. The average flux during the non-eclipse phase should be near or higher than the average flux across all observations. To improve the signal-to-noise ratio, observations within the potential eclipse phase are stacked and compared against the stacked spectrum from the non-eclipse phase. Based on the following criteria, we further select high-confidence eclipse events:

(3) Criterion 3: The stacked spectrum within the potential eclipse phase shows significant differences compared to the non-eclipse phase, and this variation can be well fitted by the partial covering absorption (PCA) model. The spectral fitting procedure is described in the following section.

(4) Criterion 4: During the eclipse phase, at least three consecutive points from the time-resolved spectral fitting must yield well-constrained $N_{\text{H}}$ and covering factor (CF) values in the PCA model, with these values being at least 3$\sigma$ higher than those outside the eclipse phase. The time-resolved analysis uses spectra stacked within several-day time bins (except for NGC 3783; see Section \ref{sec:3.2} for details). Additionally, the $N_{\text{H}}$ and CF light curves should exhibit at least one of the ingress or egress transition processes.

Eventually, a high-confidence eclipse event should satisfy all four criteria above. Additionally, some events do show a significant change in spectral shape during the eclipse phase, satisfying the first three criteria. However, the lack of sampling leads to the difficulty of constraining the eclipse duration. Since this does not pass Criterion 4, we define it as a candidate event. Some other candidates do not satisfy Criterion 1 as their hardness ratios do not vary so significantly, but their flux variations are similar to those in an eclipse event. We report some typical events with good XRT sampling and define them as candidate eclipse events.

\subsection{Analysis of the Eclipse and Non-eclipse Spectra} 
\label{sec:3.2}

In this work, we conduct model fitting for the defined potential eclipse and non-eclipse phases. For each source, the selection of these two phases is performed by visual inspection (see Section \ref{sec:3.1} for definition). To better constrain the intrinsic properties of the host AGNs, we also ensure that the stacked spectrum from XRT observations during the non-eclipse phase has a high quality (i.e., at least 1800 counts). All of our spectral analyses are conducted using \texttt{XSPEC} version \texttt{12.13.1} \citep{1996ASPC..101...17A}.  We limit our spectral fitting to the 0.3--6 keV band because the higher energy band might include additional components such as iron K$\alpha$ lines and reflection, while the signal-to-noise of XRT spectra is generally not enough to distinguish these components. 

In the non-eclipse phase, the stacked spectrum is fitted using a model comprising \texttt{compTT} \citep{1994ApJ...434..570T} and \texttt{nthComp} \citep{1996MNRAS.283..193Z,1999MNRAS.309..561Z} in \texttt{XSPEC}, with the Galactic absorption modeled by \texttt{TBabs} \citep{2000ApJ...542..914W}. The $N_{\text {H,gal}}$ is fixed at the Galactic value along the source’s line of sight \citep{2013MNRAS.431..394W}. In this model scenario, \texttt{compTT} is used to account for the soft X-ray excess, assuming it originates from the comptonization in a warm corona \citep[e.g.][]{1987ApJ...321..305C} with seed photons supplied by the accretion disc. The \texttt{nthComp} component models the X-rays above 2 keV, which are assumed to be produced by the comptonization in a hot corona \citep[e.g.][]{1993ApJ...413..507H,1995ApJ...438L..63Z} with an electron temperature fixed at 200 keV.

Based on the best-fit spectral model for the non-eclipse phase, we multiply a PCA \citep[zxipcf,][]{2008MNRAS.385L.108R} component to fit the eclipse spectrum. Since the source flux is low during the eclipse, to enhance the spectral quality, we first stack the XRT spectra within the potential eclipse phase, and fit the stacked spectrum to obtain the values of $N_{\text H}$, CF, and the ionization parameter log$\xi$. These values represent the mean properties of the obscurer during the eclipse.

\begin{figure}[ht!]
\centering
\includegraphics[width=0.45\textwidth]{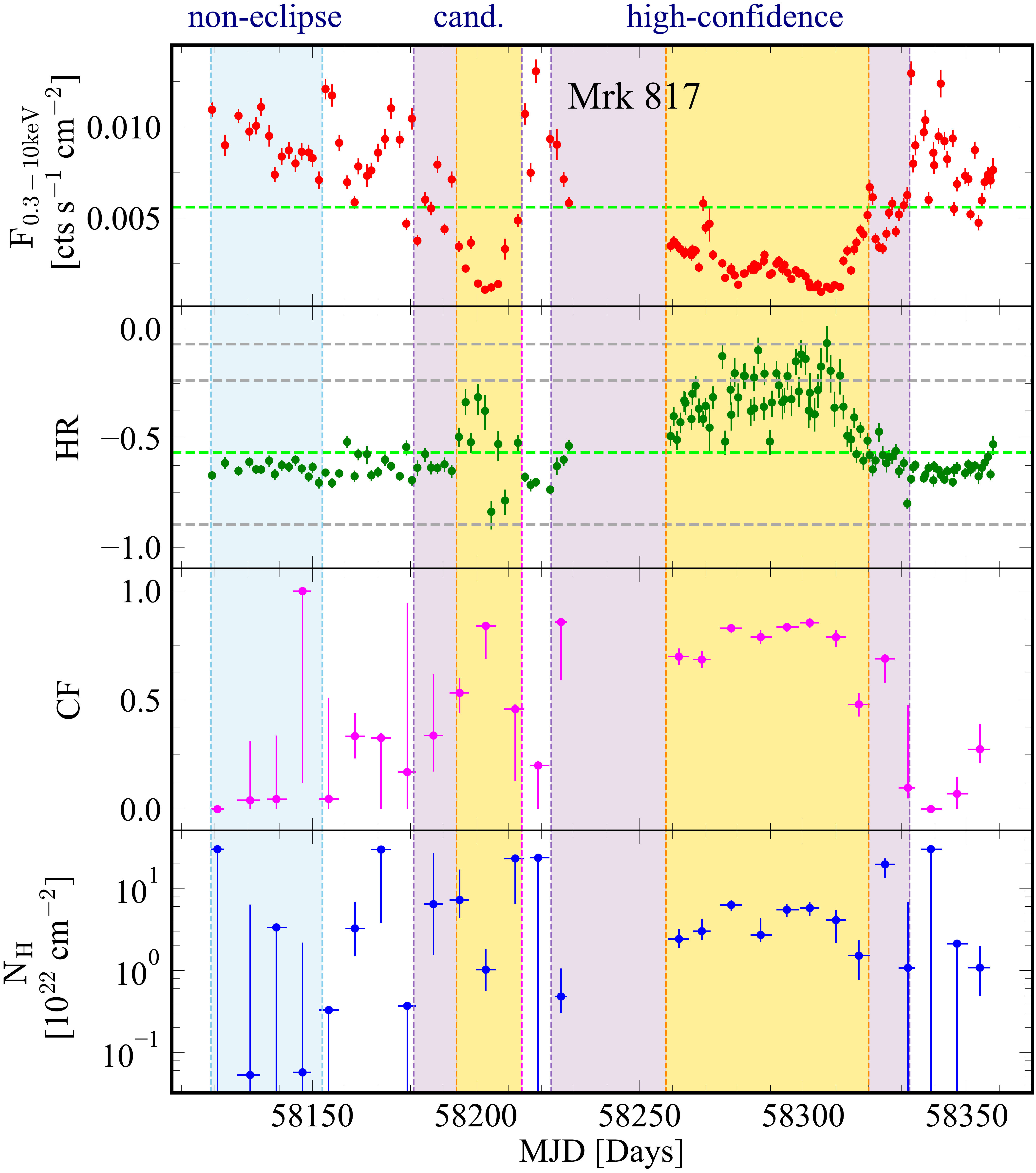}
\caption{The X-ray, hardness ratio, and time-resolved spectrum fitting parameters ($N_{\text H}$ and CF with 1$\sigma$ uncertainties) light curves of Mrk 817. The time-resolved spectral fitting is performed using a 7-day bin. The lime dashed lines stand for the average value of flux and hardness ratio. In panels 1 and 2, the grey dashed lines located near to far from the lime dashed line represent the 2$\sigma$ and 3$\sigma$ standard deviations. The blue-shaded region represents the non-eclipse phase, and the lower limits of eclipse duration are highlighted in gold regions. The combination of these gold-highlighted regions and the light purple regions denotes the upper limit. The upper limit of eclipse duration of the candidate event Mrk 817/2018.3 is from MJD 58180.4 to 58215.0, while the high-confidence event Mrk 817/2018.5 occurs during MJD 58222.7 - 58333.0.
\label{fig:1}}
\end{figure}

Then, to study the variation of these parameters during eclipses, we perform a time-resolved spectra analysis. Observations from eclipse-occurring periods, combined with the defined non-eclipse phase for comparison, are binned into several-day intervals, and then the spectra within each bin are stacked. To balance time resolution and spectral quality, we test different bin sizes for each eclipse event and select the optimal one (the time spans of each bin are indicated by horizontal error bars in the $N_{\rm H}$ and CF light curves, e.g., see Fig. \ref{fig:1}). As a result, each source may have different time bins, depending on their brightness and detection cadence (except for NGC 3783, since it is bright enough and has limited eclipse-period observations, we do not bin and stack the spectra to maintain time resolution). Given the lower quality of these spectra, we primarily focus on the variation of $N_{\text H}$ and CF, while the log$\xi$ is fixed at the average value. Additionally, we assume that the change of spectral shape during the eclipse is primarily due to the emergence of an obscurer. Therefore, we keep the intrinsic spectral shape fixed to that of the non-eclipse phase, allowing only the overall normalization to vary. This is implemented by multiplying an additional constant to the model. The final \texttt{XSPEC} model for the eclipse period is \texttt{TBabs*zxipcf*constant*(compTT + nthComp)}, and only the $N_{\text H}$, CF parameters of \texttt{zxipcf}, and the normalization constant are allowed to vary. Using this model, we measure the variation of $N_{\text H}$ and CF during the eclipse phase of each source.

\begin{deluxetable*}{CCCCCCCCC}
\tablenum{1}
\renewcommand{\arraystretch}{1.1}
\tablecaption{Results of eclipse events 
\label{tab:1}}
\tablehead{
\colhead{Source Name} & \colhead{Type}& \colhead{Event$^a$} & \colhead{Duration$^b$} & \colhead{$N_{\text H}$ $^c$} & \colhead{CF$^c$} & \colhead{log $\xi$ $^c$} & \colhead{$\chi^{2}_{red}/d.o.f$} & \colhead{Comments} \\
\colhead{} & \colhead{} & \colhead{} & \colhead{(d)} & \colhead{(10$^{22}$ cm$^{-2}$)} & \colhead{} & \colhead{(erg s$^{-1}$ cm)} & \colhead{} & \colhead{}
}
\startdata
\multicolumn{9}{c}{High-confidence Event} \\
\textnormal{Mrk 817}    &	\textnormal{Sy1.5}  &	2018.5 	&	58.9$-$110.3   &	5.45$^{+0.42}_{-0.35}$	&	0.825$^{+0.006}_{-0.006}$	&	0.708$^{+0.214}_{-0.167}$	&  1.33/86  &  \\
\textnormal{1H 0707-495}&	\textnormal{NLS1}	&	2010.11 &  84.2$-$164.0 	&
23.43$^{+7.72}_{-1.05}$	&	0.877$^{+0.017}_{-0.018}$	&	1.936$^{+0.023}_{-0.024}$	&  0.94/42 \\
\textnormal{Mrk 509}	&	\textnormal{Sy1.2}	&	2017.5	& 63.5$-$83.0	&	
0.19$^{+0.05}_{-0.03}$	&	0.595$^{+0.069}_{-0.068}$	&	-0.547$^{+0.011}_{-0.122}$	&  1.27/258\\
\multicolumn{9}{c}{Candidate Event} \\
\textnormal{Mrk 817}	&	\textnormal{Sy1.5}	&	2018.3 	&	18.0$-$34.6 	&	5.33$^{+1.65}_{-0.91}$	&	0.598$^{+0.035}_{-0.023}$	&	0.903$^{+0.331}_{-0.436}$	&  0.85/67  & \textnormal{fail criterion 1}\\
\textnormal{NGC 6814}	&	\textnormal{Sy1.5}	&	2016.3.30 	&  $>$1.6	&	4.52$^{+1.05}_{-0.90}$	&	0.863$^{+0.042}_{-0.049}$	& 	1.485$^{+0.154}_{-0.153}$	&  0.69/40 & \textnormal{fail criterion 4}\\
&	& 2016.4 & $>$ 3.5 &	3.53$^{+0.84}_{-0.67}$	&	1.000$_{-0.121}$	&	2.102$^{+0.075}_{-0.128}$	&  1.02/79  & \textnormal{fail criterion 4}\\
\textnormal{NGC 3227}	&	\textnormal{Sy1.5}	&	2008.10 	& $>$ 35.3	&	14.78$^{+1.00}_{-1.03}$	&	0.838$^{+0.016}_{-0.016}$	&	1.966$^{+0.029}_{-0.030}$	&  1.19/183 & \textnormal{fail criterion 4}\\
&	&	2019.10 	& 51.6$-$324.0 &	4.72$^{+0.93}_{-0.83}$	&	0.710$^{+0.018}_{-0.018}$	&	0.754$^{+0.383}_{-0.589}$	&  0.84/94 & \textnormal{fail criterion 1 \& 4}\\
\textnormal{NGC 3783}	&	\textnormal{Sy1}	&	2022.12 	&  $>$21.6	&	1.42$^{+0.50}_{-0.27}$	&	0.869$^{+0.013}_{-0.014}$	&	-1.118$^{+0.310}_{-0.134}$	&  1.10/92 & \textnormal{fail criterion 4} \\
\textnormal{Mrk 841}	&	\textnormal{Sy1.5}	&	2013.12.28 	&  $>$1.1 	&	3.00$^{+0.79}_{-0.38}$	&	0.918$^{+0.031}_{-0.027}$	&	0.432$^{+0.346}_{-0.804}$	&  0.83/27  &  \textnormal{fail criterion 4}\\
\textnormal{MR 2251-178}	&	\textnormal{Sy1.5}	&	2020.10 	&  $-$  &	0.52$^{+0.05}_{-0.05}$	&	 0.752$^{+0.014}_{-0.014}$	& 	-0.572$^{+0.041}_{-0.117}$	&  1.22/165  &  \textnormal{fail criterion 4}\\
\enddata
\tablecomments{$^a$The estimated starting time (UTC) of the events. In NGC 6814, NGC 3227, Mrk 841 and MR 2251-178, due to limited observational sampling, we are only able to estimate the peak eclipse time. $^b$The duration of the eclipse event. The lower and upper limits are determined by the flux dropping/returning to the average value of total light curves and non-eclipse phases, respectively. For some candidate events, due to insufficient sampling, we cannot estimate the duration of the eclipse or can only give a lower limit. $^c$The results are derived by fitting the stacked spectra of the eclipse phase, while the intrinsic parameters are fixed based on the fit to the non-eclipse phase spectra. The errors on these parameters are within the 1$\sigma$ level, and the spectra with model fitting are shown in Section \ref{sec:results} for high-confidence eclipse events and Appendix \ref{app:B} for candidate eclipse events.
}
\end{deluxetable*}

\section{Results}
\label{sec:results}

Based on the above strategy, we successfully identify 3 high-confidence eclipse events in 3 sources. For each event, we define a lower limit on the eclipse period as the duration during which the flux remained below the average flux of the entire light curve. The upper limit is determined by the interval between the flux dropping below and recovering above the non-eclipse phase average flux, marking the start and end of the event. Additionally, we find 8 candidate events in 6 sources. These candidate events either satisfy only part of the criteria aforementioned or have insufficient observational sampling, making it much more difficult to constrain the duration of the eclipse periods accurately. We present the high-confidence eclipse events below (the details of candidate events are shown in Appendix \ref{app:B}). The fitting results for the eclipse clouds are summarized in Table \ref{tab:1}. The intrinsic parameters derived from the stacked spectra in the non-eclipse phase are shown in Table \ref{tab:appB}, while time-resolved spectral analysis results are listed in Table \ref{tab:appC}.

\begin{figure}[ht!]
\centering
\includegraphics[width=0.45\textwidth]{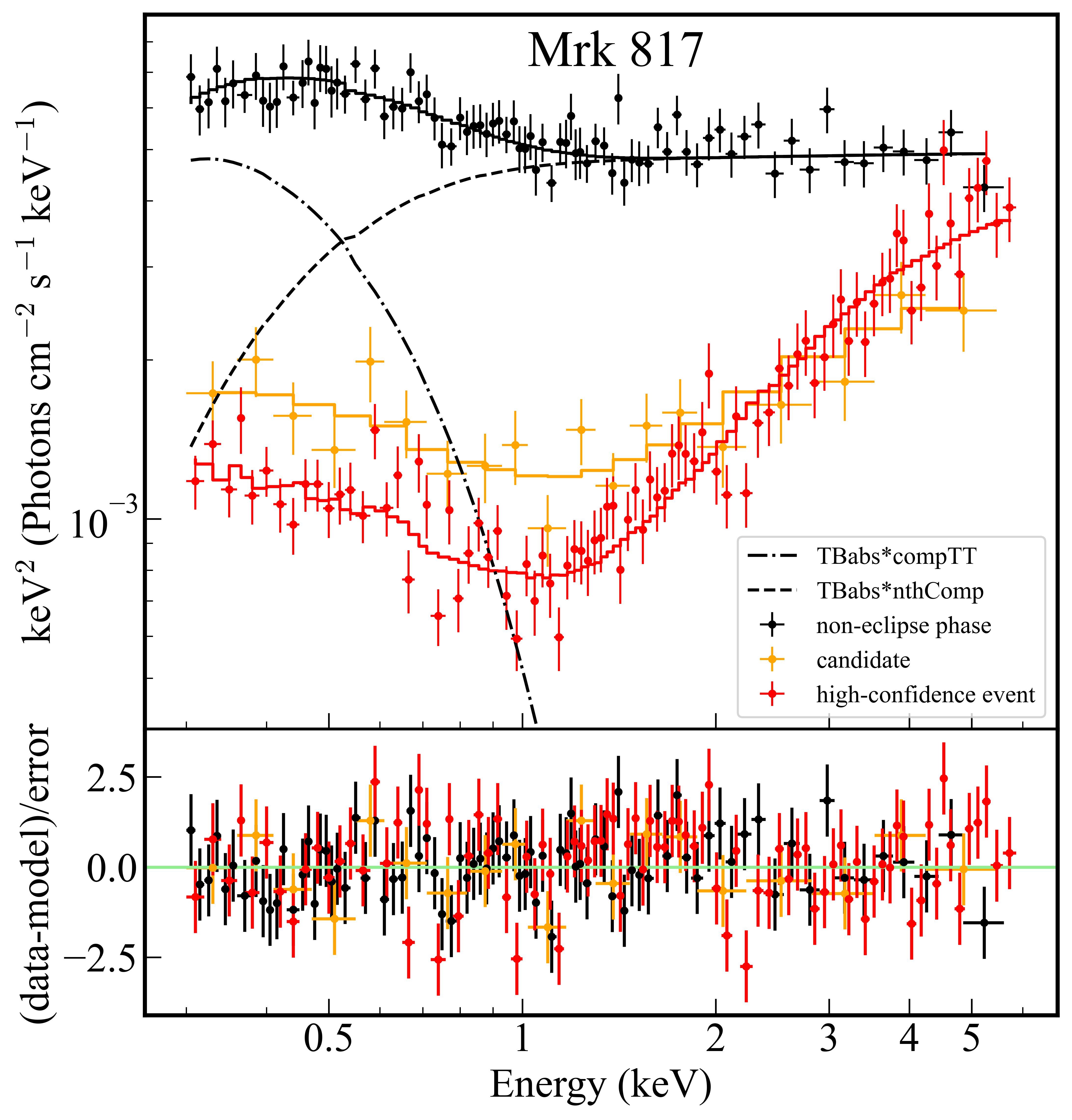}
\caption{The best fitting results of Mrk 817 during the non-eclipse and eclipse phases. For the spectra of the eclipse event, the intrinsic parameters are fixed at the same value with non-eclipse spectrum results (the value of parameters are listed in Table \ref{tab:1} and Table \ref{tab:appB}). The time durations of the non-eclipse phase and identified events are highlighted in Fig. \ref{fig:1}. 
\label{fig:2}}
\end{figure}

\subsection{Mrk 817} 
Mrk 817 is a Seyfert 1.5 galaxy at a redshift of 0.031. With $M_{\rm BH}$ of $\sim 4 \times 10^7$ $M_\odot$ \citep{2015PASP..127...67B}, it is accreting at an Eddington ratio $\sim$ 0.2 \citep{2021ApJ...922..151K}. Through a multi-wavelength intensive monitoring campaign that began in late 2020, Mrk 817 shows a strong obscuration feature both in X-ray and UV spectra, which can be attributed to a partially ionized absorber in the inner BLR \citep{2021ApJ...922..151K}. This obscurer is associated with a disk wind that disrupts reprocessing and leads to the poor correlation between X-ray and UV variation \citep{2021ApJ...911L..12M}, has been further extensively studied by \citet{2023ApJ...947....2P,2024ApJ...972..141D,2024ApJ...974...91Z}. 

Fig. \ref{fig:1} shows the light curves of Mrk 817 and the corresponding 7-day binned spectral fitting results, where at least one high-confidence eclipse event is identified. The figure also shows the non-eclipse phase we used as a reference. During the eclipse period, the flux significantly decreases, and the hardness ratio rises noticeably, which are hallmark features of an eclipse.

Fig. \ref{fig:2} presents the stacked spectra for both the eclipse and non-eclipse phases. The double-Comptonization model fits the non-eclipse spectrum well, while the eclipse spectrum shows significant changes in shape, which can be effectively modeled by including an additional PCA component. As shown in Table \ref{tab:1}, the average parameters for the high-confidence eclipse period are CF = 0.83, $N_{\text{H}} = 5.5 \times 10^{22}$ cm$^{-2}$, and log $\xi$ = 0.71. These results indicate that the absorber responsible for the eclipse is largely covering the source, has a moderate column density, and is close to neutral. Fig. \ref{fig:1} also shows the detailed variations of CF and $N_{\text{H}}$ during the eclipse. While these parameters cannot be well constrained during the non-eclipse period due to the absence/insignificance of a partial covering component, their substantial values with significant variations during the eclipse provide strong evidence for the emergence of an additional absorber.

Additionally, we identify one candidate event in the light curves of Mrk 817. This event is shorter in duration and also exhibits a noticeable increase in hardness ratio but does not satisfy Criterion 1 described in the previous section, leading us to classify it as a candidate. Fig. \ref{fig:2} also displays the stacked spectrum for this candidate event, showing a spectral shape similar to that of the high-confidence event. The fitting results in Table \ref{tab:1} reveal comparable ionization levels and column densities for both events, although the candidate event has a smaller CF of 0.60.

\begin{figure}[ht!]
\centering
\includegraphics[width=0.45\textwidth]{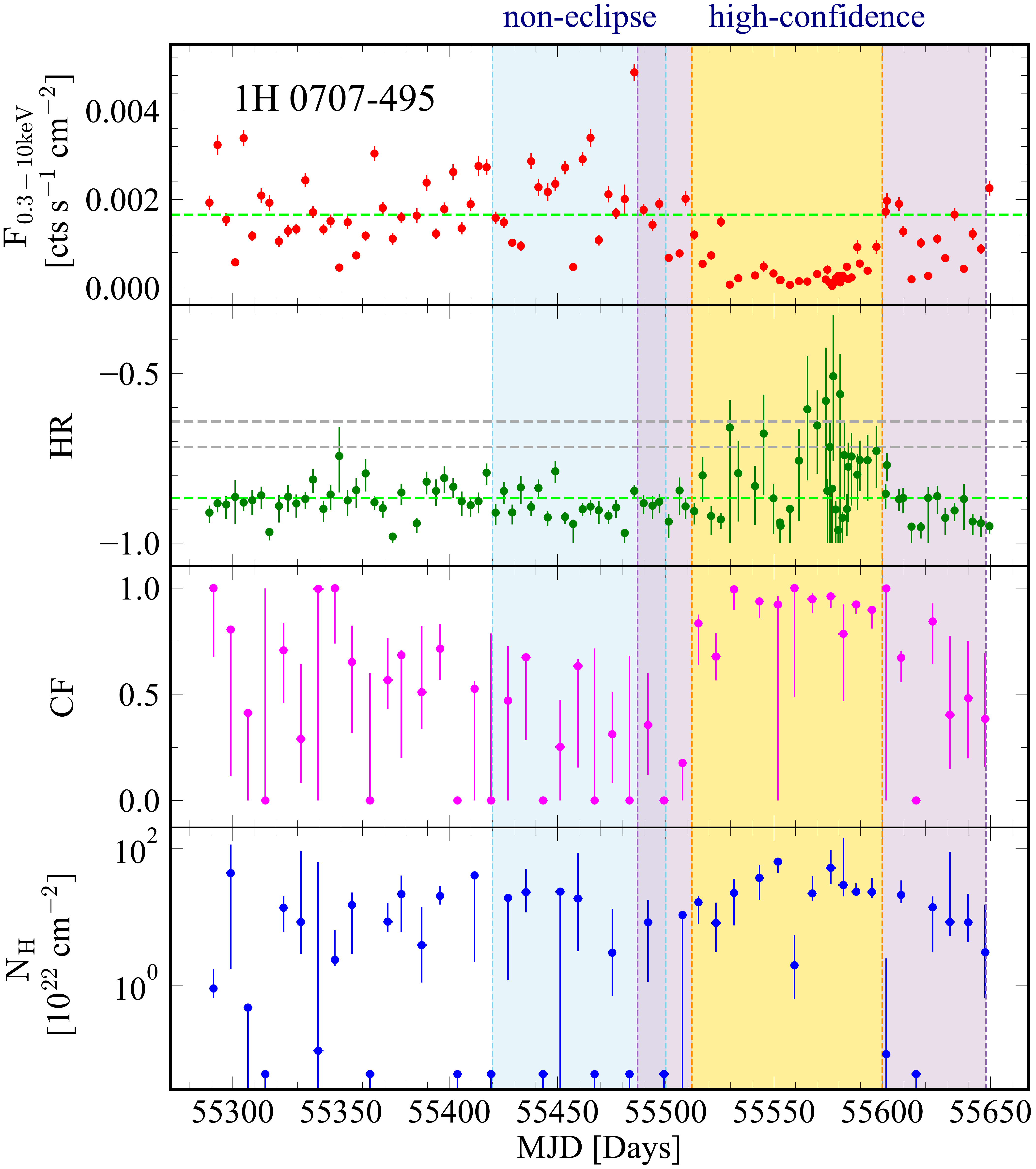}
\caption{The X-ray, hardness ratio, and time-resolved spectrum fitting parameters light curves of 1H 0707-495. The time-resolved spectral fitting is performed using a 5-day bin. The highlighted regions indicate the non-eclipse (blue) and eclipse (light purple and gold) periods. The upper limit duration of the high-confidence event 1H 0707/2010.11 is defined between the MJD 55485.5 and 55649.5.
\label{fig:3}}
\end{figure}

\begin{figure}[ht!]
\centering
\includegraphics[width=0.45\textwidth]{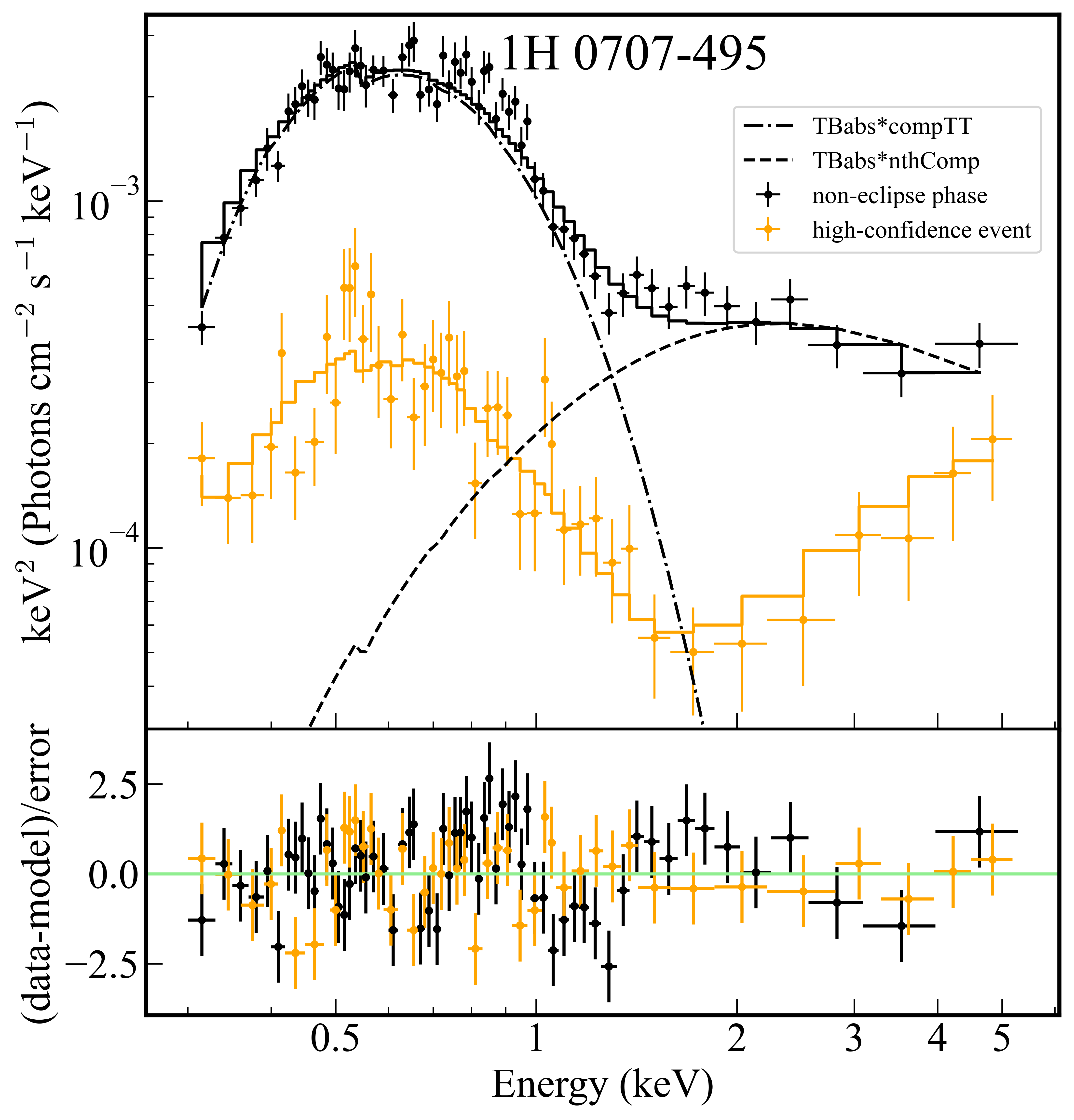}
\caption{The best fitting results of 1H 0707-495 during the non-eclipse and eclipse phases, the value of parameters are listed in Table \ref{tab:1} and Table \ref{tab:appB}.
\label{fig:4}}
\end{figure}

\subsection{1H 0707-495} 
1H 0707-495 is a bright narrow-line Seyfert 1 (NLS1) galaxy at a redshift of 0.041. It has a $M_{\rm BH}$ of $\sim 2 \times 10^6$ $M_\odot$, and an extremely super-Eddington accretion rate \citep{2016MNRAS.460.1716D}. As expected for such a system, the strong outflows detected in 2008 and 2010 provide a plausible explanation for the source's drastic X-ray variability \citep{2012MNRAS.422.1914D,2016MNRAS.461.3954H}. In October 2019, a significant drop in the ultrasoft X-ray band of this source was observed, which can be interpreted by an ionized partial absorber along the line of sight \citep{2021A&A...647A...6B}. 

We apply 5-day binning time to conduct time-resolved analysis. As shown in Fig. \ref{fig:3}, the XRT light curve reveals a distinct low-flux state lasting for almost 100 days, during which the hardness ratio undergoes significant variations. Fig. \ref{fig:4} compares the stacked spectra during the non-eclipse and eclipse phases. The spectral continuum in the non-eclipse phase is well-fitted by a double-Comptonization model, whereas the eclipse-phase spectrum requires an additional PCA component to achieve a reasonable fit. This absorber is characterized by a CF of 0.88, a column density of $N_{\text H} = 2.3 \times 10^{23}$ cm$^{-2}$, and an ionization parameter of log$\xi$ = 1.94. Moreover, residuals present in the spectral fitting for both the non-eclipse and eclipse phases suggest additional absorption features, possibly linked to persistent components within the outflow. Fig. \ref{fig:3} also demonstrates the significant increase in CF and $N_{\text H}$ during the eclipse compared to the non-eclipse phase, providing further support for the eclipse interpretation.

Moreover, the absorber observed in 1H 0707-495 differs from those found in Mrk 817, with a notably higher column density and ionization parameter. These differences are consistent with 1H 0707-495's smaller $M_{\rm BH}$ and higher mass accretion rate, as stronger and hotter outflows are expected in such conditions \citep{2023MNRAS.518.6065J}.

\begin{figure}[b]
\centering
\includegraphics[width=0.45\textwidth]{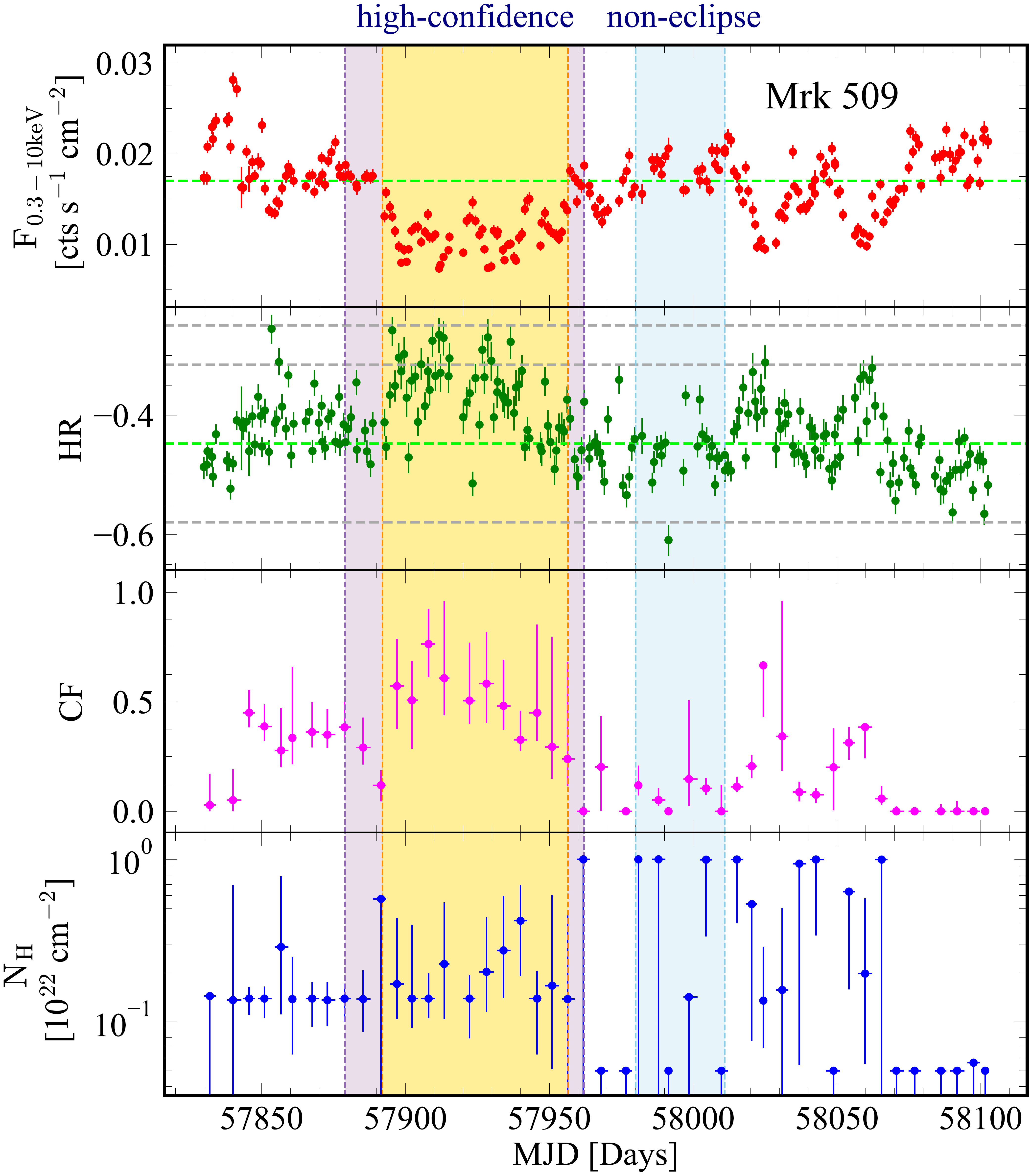}
\caption{The X-ray, hardness ratio, and time-resolved spectrum fitting parameters light curves of Mrk 509. The time-resolved spectral fitting is performed using a 5-day bin. The upper limit of high-confidence event Mrk 509/2017.5 duration occurred in MJD 57879.1 - 57962.1.
\label{fig:5}}
\end{figure}

\begin{figure}[ht!]
\centering
\includegraphics[width=0.45\textwidth]{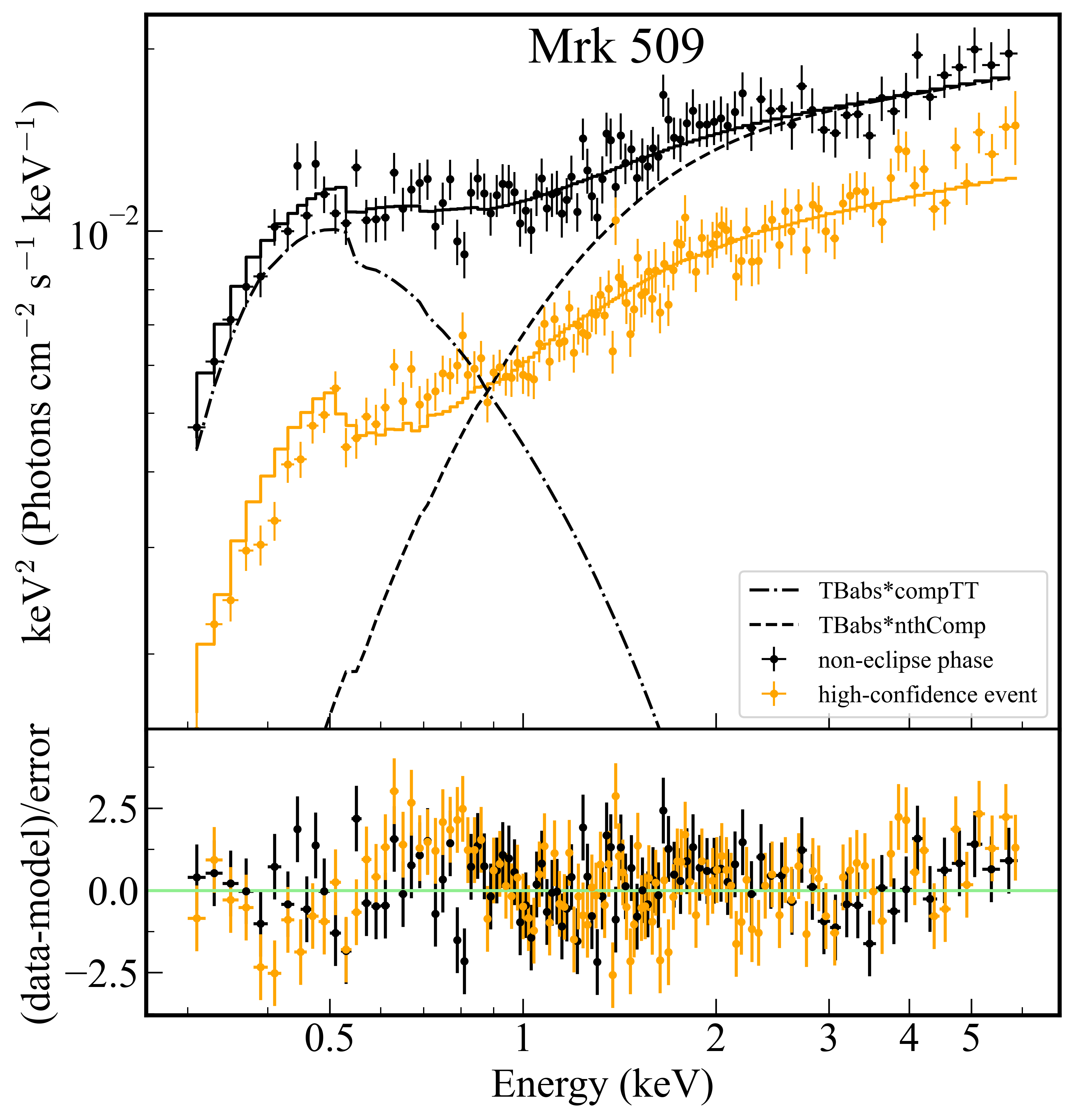}
\caption{The best fitting results of Mrk 509 during the non-eclipse and eclipse phases (see Table \ref{tab:1} and Table \ref{tab:appB} for the corresponding parameters).
\label{fig:6}}
\end{figure}

\subsection{Mrk 509}
Mrk 509 is a Seyfert 1.2 galaxy at a redshift of 0.034. Its central $M_{\rm BH}$ was $\sim 1 \times 10^8$ $M_\odot$ with an Eddington ratio of 0.16 \citep{2024A&A...684A.167G}. A soft-excess component was identified in its X-ray spectrum \citep{1985ApJ...297..633S}, and the soft X-ray absorption feature can be explained by warm absorbers \citep{2011A&A...534A..36K}, which can be associated with high-velocity outflows \citep{2009A&A...504..401C}. One eclipse event, which lasted 26 - 91 days, was detected by {\it RXTE} in 2005.9 (\citetalias{2014MNRAS.439.1403M}).

Fig. \ref{fig:5} shows an eclipse event in Mrk 509 that began around MJD 57880 and lasted for several tens of days. In Fig. \ref{fig:6}, we show that the non-eclipse stacked spectrum is well modeled by a double-Comptonization model, although the spectrum appears relatively hard, suggesting the presence of some absorption even during the non-eclipse phase. During the eclipse period, the stacked spectrum hardens further and can be adequately fitted with an additional PCA component. As summarized in Table \ref{tab:1}, the absorber during the eclipse has a CF of 0.60, a column density of $N_{\text H}=1.9 \times 10^{21}$ cm$^{-2}$, and an ionization parameter of log$\xi = - 0.55$. There was evidence of an increase in CF when the source entered the eclipse period. Moreover, the ingress phase of this eclipse lasted only about 10 days, while the egress phase extended over approximately 50 days. This asymmetry might correspond to the structure of the absorber, possibly resembling a 'comet-like' absorber as proposed by \citet{2010A&A...517A..47M} for NGC 1365.

Compared to Mrk 817 and 1H 0707-495, the absorber in Mrk 509 exhibits significantly lower column density, covering fraction and ionization. This discrepancy may be attributable to Mrk 509's larger $M_{\rm BH}$ of $1\times10^{8}~M_{\odot}$ and lower mass accretion rate, so its outflows may be weaker and cooler. We will explore this potential correlation in more detail in the next section.

Additionally, we notice in Fig. \ref{fig:5} that there are other periods where Mrk 509 exhibited ingress and egress shapes in its X-ray flux with inverse trends in the hardness ratio (e.g., MJD 58010--58080). During these periods, the covering fraction varied with lower peak CF values compared to the defined eclipsing interval. However, the peak hardness ratio during these intervals did not exceed the 2$\sigma$ threshold, and the column density showed complicated variations, so it is inconclusively identify them as eclipse events.

\begin{deluxetable*}{CCCCCCCCCC}
\renewcommand{\arraystretch}{1} 
\setlength{\tabcolsep}{2pt}
\tablenum{2}
\tablecaption{The physical properties of the eclipse clouds
\label{tab:2}}
\tablehead{
\colhead{Source Name} & \colhead{Event} & \colhead{log $\dot{m}$ $^a$} & \colhead{log $L_{\text{ion}}$,$L_{\text{bol}}$ $^a$} & \colhead{$r_{\text c}$ $^b$} & \colhead{$r_{\text c}$ $^b$} & \colhead{$r_{\text c}$/$R_{\rm d}$ $^b$} & \colhead{$D_{\text c}$ $^b$} & \colhead{$D_{\text c}$ $^b$} & \colhead{$n_{\text H}$ $^b$}  \\
 & & & \colhead{(erg s$^{-1}$)} & \colhead{(light days)} & \colhead{(10$^{4}$ $R_{\rm g}$)} & & \colhead{(light days)} & \colhead{($R_{\rm g}$)} & \colhead{(10$^{8}$ cm$^{-3}$)}
}
\startdata
\multicolumn{10}{c}{Confirmed eclipse event} \\
\textnormal{Mrk 817} & 2018.5 & $-$0.61& 45.03,45.16 & 514.6\pm109.3 & 23.23\pm4.93 & 0.90\pm0.19 & 0.175\pm0.046 & 79.13\pm20.64 & 1.311\pm0.526\\
\textnormal{1H 0707-495} & 2010.11 & 1.79 & 46.22,46.23 & 178.9\pm28.7 & 153.93\pm24.70 & 0.09\pm0.01 & 0.099\pm0.025 & 848.20\pm218.40 & 10.063\pm5.992   \\
\textnormal{Mrk 509} & 2017.5 & $-$0.72 & 44.95,45.28 & 6738.1\pm832.6 & 118.36\pm14.63 & 10.24\pm1.27 & 0.068\pm0.008 & 11.87\pm1.46 & 0.109\pm0.023 \\
\multicolumn{10}{c}{Candidate eclipse event} \\
\textnormal{Mrk 817} & 2018.3 & $-$0.61 & 45.03,45.16 & 288.7\pm112.2 & 13.04\pm5.07 & 0.50\pm0.20 & 0.075\pm0.024 & 33.97\pm10.93 & 3.131\pm1.855 \\
\textnormal{NGC 6814} & 2016.3.30 & $-$1.32 & 43.52,43.81 & $>$11.1 & $>$1.77 & $>$0.09 & $>$0.012 & $>$19.47 & $<$14.231\\
 & 2016.4 & & & $>$ 9.4 & $>$ 1.50 & $>$ 0.08 & $>$ 0.029 & $>$ 46.12 & $<$ 4.669 \\
\textnormal{NGC 3227} & 2008.10 & $-$1.59 & 43.33,43.97 & $>$ 16.3 & $>$ 0.67 & $>$ 0.11 & $>$ 0.431 & $>$ 177.64 & $<$ 1.321 \\
 & 2019.10 & & & 166.5\pm107.3 & 6.86\pm4.42 & 1.14\pm0.74 & 0.742\pm0.428 & 305.49\pm176.30 & 0.580\pm1.842 \\
 \textnormal{NGC 3783} & 2022.12 & $-$1.21 & 44.27,44.47 & $>$1365.6 & $>$81.28 & $>$5.28 & $>$0.024 & $>$14.56 & $<$2.197 \\
\textnormal{Mrk 841} & 2013.12.28 & $-$1.24 & 44.63,45.00 & $>$153.4 & $>$1.91 & $>$0.32 & $>$0.009 & $>$1.10 & $<$13.850 \\
\textnormal{MR 2251-178} & 2020.10 & $-$0.69 & 45.14,45.52 & $-$ & $-$ & $-$ & $-$ & $-$ & $-$ \\
\enddata
\tablecomments{
$^a$The best-fitting results of model \texttt{optxagnf} during the non-eclipse phases are given, with each parameter having 1$\sigma$ uncertainty. $\dot{m}$ is the mass accretion rate; $L_{\text{ion}}$ is the 1-1000 Rydberg ionizing luminosity; $L_{\text{bol}}$ is the bolometric luminosity integrated from 0.001 to 100 keV. $^b$The best-estimated value of the cloud physical parameters. The uncertainties are composed of $N_{\text H}$, $\Delta t$ and log$\xi$. For some candidate events, due to the lack of sampling, we either cannot determine their $\Delta t$ or can only give a lower limit for it. As a result, we either cannot calculate the properties of the eclipse clouds or can only estimate the lower and upper limits of the physical parameters (considering the uncertainties of $N_{\text{H}}$ and $\log\xi$).
}
\end{deluxetable*}

\section{Discussion}
\label{sec:discussion}

\subsection{Properties of the Obscuring Clouds}

\subsubsection{Size and Spatial Location}
\label{sec:5.2.1}

Eclipse events are caused by obscuring clouds temporarily crossing the line of sight to the X-ray source. Following the method of \citet{2003MNRAS.342L..41L}, we can constrain the distance ($r_{\rm{c}}$) between the obscuring clouds and the central black hole. We simplify the model by assuming that each cloud has uniform density and ionization state and moves in a Keplerian orbit. The ionization parameter $\xi$ is defined as: $\xi = L_{\rm{ion}}/(n_{\rm H}r_{\rm c}^2)$, where  $L_{\text{ion}}$  is the ionizing luminosity estimated in the energy range of 13.6 eV to 13.6 keV, $n_{\text{H}}$ is the hydrogen number density of eclipse cloud, and $r_{\text{c}}$ is the distance between the cloud and the central X-ray source. The diameter of the cloud $D_{\text{c}}$ can be expressed as: $D_{\text{c}} =\nu_{\text{cloud}} \Delta t$, where $\nu_{\text{cloud}}$ is the cloud's Keplerian velocity, and $\Delta t$ is the duration of the eclipse event. The Keplerian velocity is given by: $\nu_{\text{cloud}} = \sqrt{GM_{\text{BH}}/r_{\text c}}$, where $M_{\rm{BH}}$ is the central black hole mass, and $G$ is the gravitational constant. Combining these equations, the distance of the cloud can be determined as: $r_{\text c} \sim$ ($GM_{\rm{BH}}$)$^{1/5}$ $L_{\text{ion}}^{2/5}$ $\Delta t^{2/5}$ $N_{\text H}^{-2/5}$ $\xi^{-2/5}$. This formula allows us to estimate the location of the obscuring clouds relative to the central black hole based on observable parameters such as $L_{\rm{ion}} $, $N_{\rm{H}}$, $\xi$ and $\Delta t$.

The estimation of $L_{\text{ion}}$ is model-dependent and could be strongly affected by the absorption component during the eclipse phase. We assume that the Spectral Energy Distribution (SED) of the central ionizing source seen by the obscuring clouds is similar to the SED observed during the non-eclipse phase. To reconstruct the intrinsic broadband SED, we use the {\it Swift}/UVOT and XRT data from the non-eclipse phase. The data are fitted using the \texttt{optxagnf} model \citep{2012MNRAS.420.1848D} in \texttt{XSPEC}, which comprises the accretion disc emission and Comptonized emission in AGNs. We fix the $M_{\rm BH}$ from previous studies (listed in Table \ref{app:A}), and add the absorption (\texttt{TBabs}) and reddening \citep[\texttt{zredden}, IR/optical/UV extinction from][]{1989ApJ...345..245C} component when needed. The best-fit mass accretion rate and the derived luminosity are listed in Table \ref{tab:2}.

To better constrain the physical parameters of the eclipse clouds, we perform 1000 random samplings of key parameters ($\Delta t$, $N_{\text H}$, and $\xi$) within their respective error ranges, and then calculate the average and standard deviation of $r_{\text c}$. The diameter of the eclipse cloud ($D_{\text c}$) is derived from $r_{\text c}$. In addition, due to non-uniform temporal sampling, we cannot determine the eclipse duration of MR 2251-178/2020.10 and calculate the physical properties of its eclipse cloud. For some candidates that only have lower limit timescales, we can only derive the lower limit of $r_{\text c}$ and $D_{\text c}$. As shown in Table \ref{tab:2}, for the 5 eclipse events with well-constrained duration, the obscuring clouds have sizes ranging from approximately 0.05 to 1 light-days. This is consistent with our method of searching for eclipses in the long-term XRT monitoring, which predominantly identifies clouds causing eclipses lasting tens to hundreds of days.

By assuming they are spherical and have a uniform gas density, we can also estimate the number density of each cloud: $n_{\rm H} = N_{\rm H}/D_{\text{c}}$. Table \ref{tab:2} lists the estimated gas density for each obscuring cloud based on this method (except MR 2251-178/2020.10), ranging from $10^{6-9}$ cm$^{-3}$. This is several orders of magnitude lower than the typical density of the BLR, which is typically $10^{9-13}$ cm$^{-3}$ \citep{2006LNP...693...77P}. However, this result does not necessarily rule out a connection to the BLR, because the actual clouds may not be in Keplerian motion or spherical. Instead, these clouds may also originate from material in outflows. Then the eclipse period may not correspond to the physical size of the cloud but rather to the duration of the outflows intervening the line-of-sight. For example, recent simulations by \citet{2024MNRAS.530.5143D} show that AGN outflows can exhibit episodic and non-steady behavior. Therefore, the density estimates provided here should have a large uncertainty.

\begin{figure}[ht!]
\centering
\includegraphics[width=0.45\textwidth]{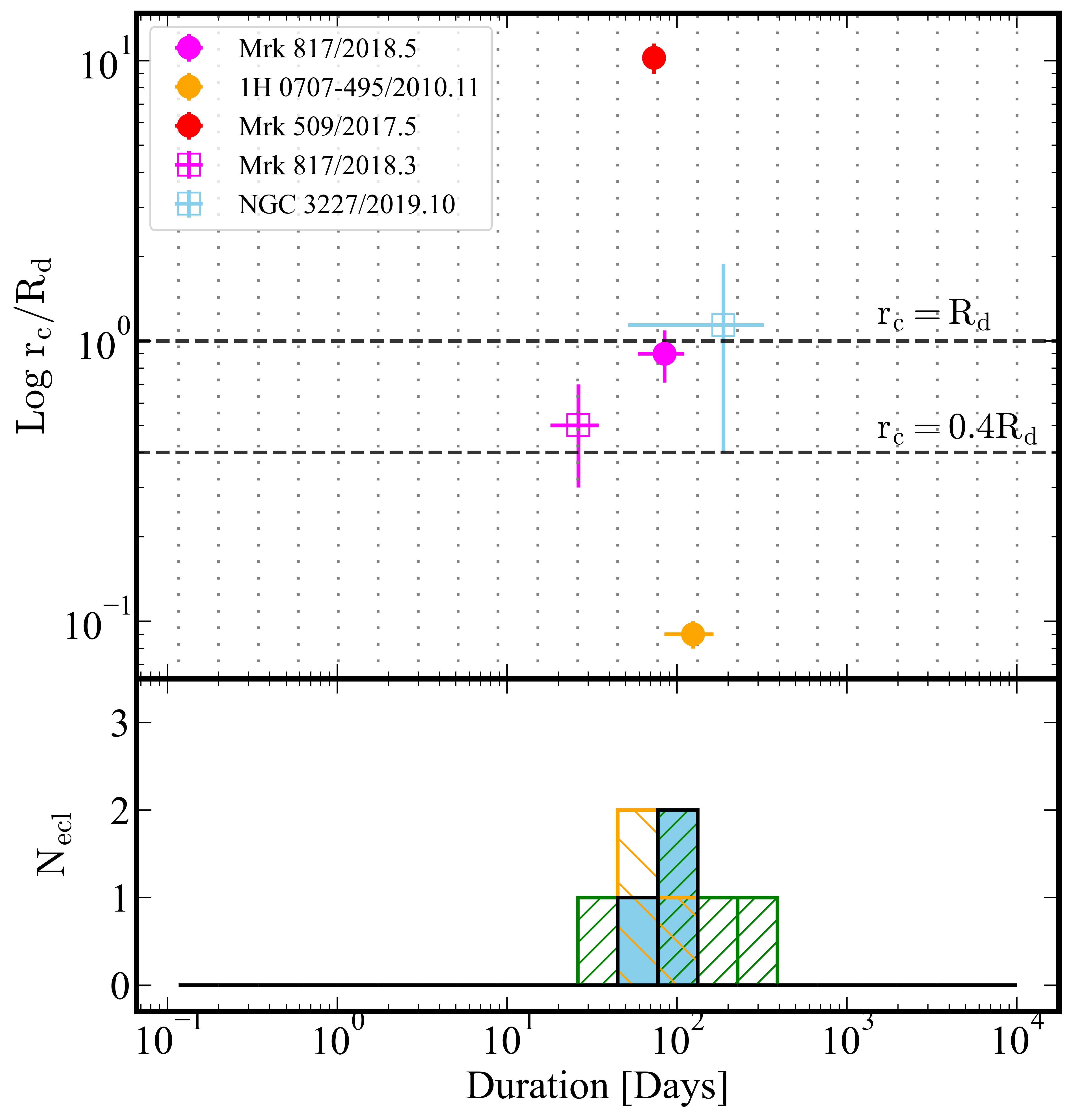}
\caption{The upper panel displays the timescale of each eclipse event and their corresponding distance ratio $r_{\text c}$/$R_{\text c}$. High-confidence events are depicted as colored dots, while candidate events are represented by square points. We only display 5 events for which the upper and lower limits of timescales can be well constrained. Moreover, the two black dashed lines denote the region of DSZ. The number of observed eclipse events for type I AGN in different timescales within the corresponding duration bins are shown in the lower panel. The best-estimated number of eclipse events $N_{\text{ecl}}$(t$_{\text i}$) is depicted as blue bars, while the minimum and maximum likelihood $N_{\text{ecl}}$(t$_{\text i}$) are indicated by orange and green solid lines, respectively. 
\label{fig:fun1}}
\end{figure}

Furthermore, we find these clouds are located at distances of a few to several thousands of light-days from the central black hole, corresponding to $2000 - 10^{6}~R_{\text g}$ where $R_{\text g}$ is the Gravitational radius defined as $GM_{\text{BH}}/c^2$. Then we compared $r_{\rm c}$ to the outer radius of the DSZ denoted as $R_{\text{d}}$, which is estimated as $R_{\text{d}} \sim 0.4~(L_{\text{ion}}/{10^{45}\text{erg s}^{-1}})^{0.5} ~(T_{\text{d}}/{1500\text{K}})^{-2.6}$\text{pc} , where $L_{\text{ion}}$ is provided in Table \ref{tab:2}, and we assume a dust temperature $T_{\text{d}} = 1500$ \text{K} \citep{2008ApJ...685..160N}. Additionally, we consider the DSZ to range from 0.4 $R_{\text{d}}$ to 1 $R_{\text{d}}$ (\citetalias{2014MNRAS.439.1403M}), while the inner regions are dust-free and the outer regions do not undergo sublimation. For each eclipse event, the ratio between $r_{\text{c}}$ and $R_{\text{d}}$ is listed in Table \ref{tab:2}. Fig. \ref{fig:fun1} presents these ratios of the five temporally well-constrained events and their eclipse durations. We find that some clouds are located within the inner radius of DSZ and exhibit higher ionization (e.g. 1H0707-495/2010.11), while others are outside the DSZ and have lower ionization (e.g. Mrk509/2017.5). As discussed in the next section, this could be related to the AGN's key parameters such as the $M_{\rm BH}$ and mass accretion rate ($\dot{m}$).

\begin{figure*}[ht!]
\centering
\includegraphics[scale=0.52]{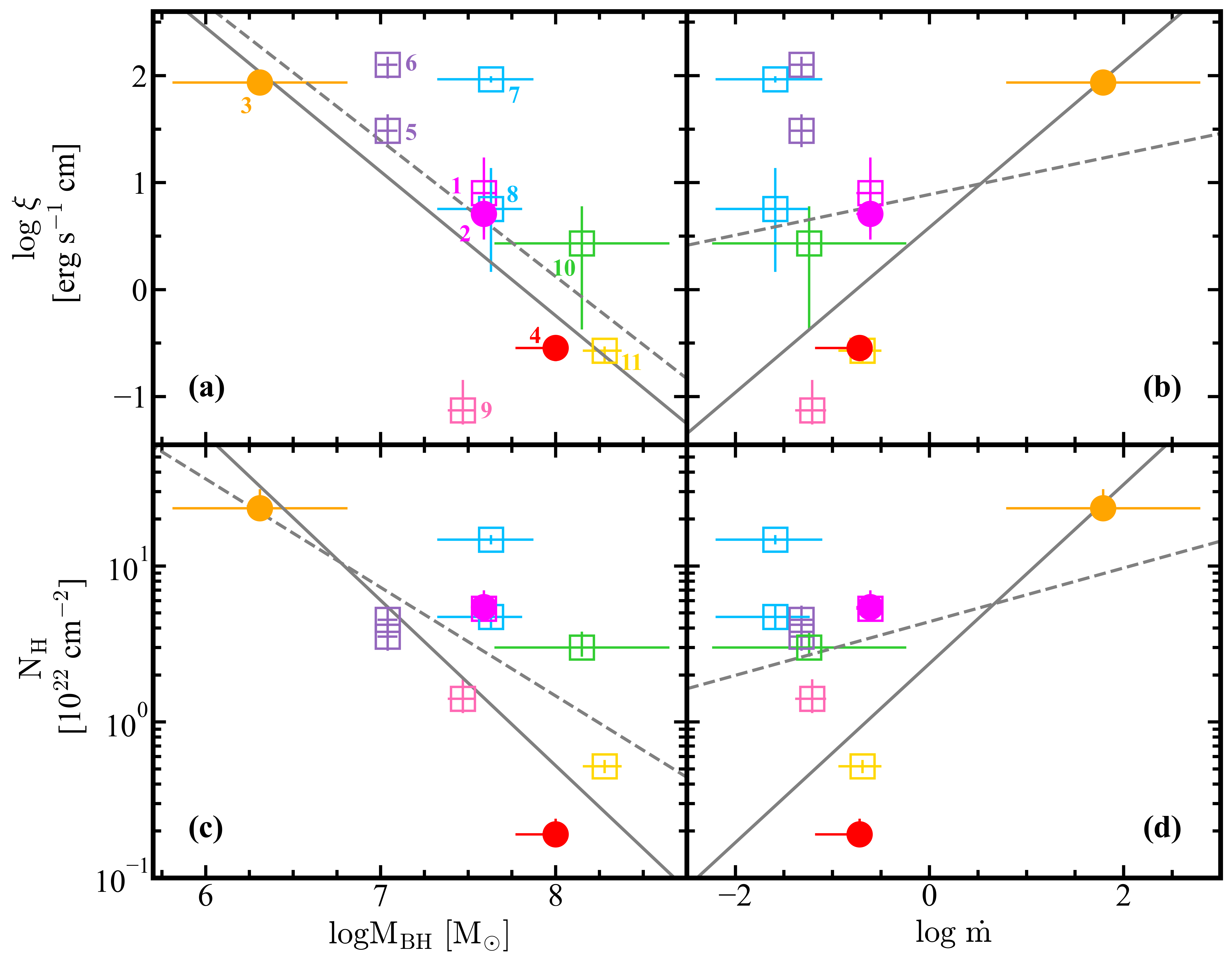}
\caption{Ionization parameter (log$\xi$) and column density ($N_{\rm H}$) of the eclipse clouds derived by fitting the stacked spectra, versus the AGN black hole mass ($M_{\rm BH}$) and mass accretion rate ($\dot{m}$). Circles and squares denote high-confidence and candidate events respectively. Eclipse clouds in the same sources are plotted in the same color. The errors in $M_{\rm BH}$ and $\dot{m}$ are complied from previous studies (see Table \ref{tab:appA}). Linear regression results based on the high-confidence events are shown as grey solid lines, and those based on all the identified events are shown as grey dashed lines. Table \ref{tab:3} listed the correlation strength between cloud properties and AGNs' key parameters. The numbers near the points indicate different identified events: 1. Mrk 817/2018.3; 2. Mrk 817/2018.5; 3. 1H 0707-495/2010.11; 4. Mrk 509/2017.5; 5. NGC 6814/2016.3.30; 6. NGC 6814/2016.4; 7. NGC 3227/2008.10; 8. NGC 3227/2019.10; 9. NGC 3783/2022.12; 10. Mrk 841/2013.12.28; 11. MR 2251-178/2020.10.
\label{fig:comp}}
\end{figure*}

\subsubsection{Dependence of Cloud Properties on $M_{\rm BH}$ and $\dot{m}$}
\label{sec:5.2.2}

The occurrence of obscuring clouds is often associated with clumps in outflows or the torus, and their observation is likely influenced by the inclination angle of the AGN. The events reported in this work are all identified in type I AGN, where the inclination angle is expected to be relatively low. Therefore, the clouds found in this sample are less likely to originate from clumps within the dusty torus. Instead, another possible origin would be the clumps in outflows, whose physical properties should be connected to the properties of the accretion flow.

Since the black hole mass ($M_{\rm BH}$) and mass accretion rate ($\dot{m}$) primarily determine the multiwavelength properties of AGN \citep[e.g.,][]{2009MNRAS.392.1124V,2011ApJ...728...98D,2012MNRAS.420.1848D,2012MNRAS.425..907J,2023MNRAS.518.6065J}, it is reasonable to expect that the properties of these clouds are also correlated with these two parameters. 
We investigate the dependence of the cloud ionization parameter (log $\xi$) and column density ($N_{\rm H}$) on $M_{\rm BH}$ and $\dot{m}$, as shown in Fig. \ref{fig:comp}. The errors in $M_{\rm BH}$ are compiled from previous studies (see Table \ref{tab:appA} for reference), while the uncertainty in $\dot{m}$ is primarily influenced by the error in $M_{\rm BH}$, with a propagated effect of square amplification \citep[ $\dot{m} \propto M_{\rm BH}^{-2}$ for an observed optical/UV luminosity, ][]{2023MNRAS.518.6065J}. Considering the uncertainty ranges of these four parameters, we compute the correlation coefficients and their associated errors, which are listed in Table \ref{tab:3}.

As described in the previous section, we find tentative correlations between cloud properties and $M_{\rm BH}$. In AGNs with smaller $M_{\rm BH}$ (e.g., 1H0707-495), the obscuring clouds tend to have higher ionization states, which can be understood as the result of higher-temperature radiation fields \citep[$T^4_{\rm eff} \propto M^{-1}_{\rm BH} \dot{m}$, ][]{1973A&A....24..337S} producing more ionizing photons to sustain elevated ionization levels. Additionally, the column density appears to be higher in low-mass systems, consistent with the fact that these systems typically exhibit higher Eddington ratios \citep[e.g.][]{2012MNRAS.425..907J}, thereby leading to stronger outflows \citep[e.g.][]{2023MNRAS.518.6065J}. However, due to the large uncertainties in the mass accretion rate measurements, no clear correlation is found between cloud properties and mass accretion rate across all identified events. Although strong positive correlations are observed in the three high-confidence events, no definitive conclusions can be drawn from this small eclipse sample.

We also emphasize that, due to the limited sample size, the aforementioned correlations are only tentative. Nevertheless, the better-defined correlations may partly result from the capability of \textit{Swift}/XRT to probe soft X-rays down to 0.3 keV, enabling better constraints on the parameters of obscuring clouds and better reflecting these underlying correlations.

\begin{deluxetable}{CCC}
\renewcommand{\arraystretch}{1} 
\setlength{\tabcolsep}{3pt}
\tablenum{3}
\tablecaption{The correlation coefficients of clouds properties and AGNs' key parameters, along with their associated uncertainties.
\label{tab:3}}
\tablehead{
\colhead{Pair} & \colhead{Pearson-r} & \colhead{Spearman-$\rho$}
}
\startdata
\multicolumn{3}{c}{11 identified events} \\
\colhead{log $M_{\rm BH}$ vs log $\xi$} & -0.645\pm0.079 & -0.554\pm0.101 \\
\colhead{log $M_{\rm BH}$ vs log $N_{\rm H}$} & -0.652\pm0.097 & -0.490\pm0.064 \\
\colhead{log $\dot{m}$ vs log $\xi$} & 0.162\pm0.090 & -0.270\pm0.110 \\
\colhead{log $\dot{m}$ vs log $N_{\rm H}$} & 0.271\pm0.107 & 0.105\pm0.083 \\
\multicolumn{3}{c}{3 high-confidence events} \\
\colhead{log $M_{\rm BH}$ vs log $\xi$} & -0.957\pm0.037 & -1.000\pm0.032 \\
\colhead{log $M_{\rm BH}$ vs log $N_{\rm H}$} & -0.871\pm0.054 & -1.000\pm0.025 \\
\colhead{log $\dot{m}$ vs log $\xi$} & 0.882\pm0.069 & 1.000\pm0.196  \\
\colhead{log $\dot{m}$ vs log $N_{\rm H}$} & 0.759\pm0.065 & 1.000\pm0.197 \\
\enddata
\end{deluxetable}

\subsubsection{Possible Origins: Outflows, BLR and Torus}
\label{sec:5.2.3}

The obscuring clouds identified in this study exhibit distinct physical properties across different accretion systems, suggesting that the mechanisms responsible for line-of-sight absorption may vary.

In Mrk 509 and NGC 3783, the eclipsing clouds have very low ionization parameters (-0.5 and -1.1), indicating relatively low temperatures. These clouds are likely located in the outer regions of the central engine, possibly within the torus, where they may be dusty and in orbital motion. This speculation aligns with the distance estimates in Section \ref{sec:5.2.1}, which place the obscuring cloud in Mrk 509 beyond the DSZ (see also Fig. \ref{fig:fun1}). Additionally, in Fig. \ref{fig:5}, we notice that the eclipse in Mrk 509 shows a significant rising trend of column density during the egress phase, resembling a comet-like structure, similar to what was observed in NGC 1365 \citep{2010A&A...517A..47M}. However, we emphasize that the method used for distance estimates involves several oversimplified assumptions, so it is not possible to rule out a connection between these clouds and the BLR. Moreover, UV line-driven remains a viable mechanism for triggering outflows in these systems with large black hole masses \citep[e.g., ][]{2021MNRAS.503.1442M}. Indeed, multi-wavelength monitoring of Mrk 817 has revealed strong outflows in this system \citep[e.g., ][]{2021ApJ...922..151K}. Therefore, the observed eclipse could also be related to outflowing material.

1H 0707-495 is the only source in our eclipse sample exhibiting super-Eddington accretion. Such extreme accretion can drive powerful outflows in this system, as suggested by previous studies \citep[e.g., ][]{2012MNRAS.422.1914D,2016MNRAS.461.3954H}. Thus, we consider the eclipse event in 1H 0707-495 to be more likely associated with clumpy material within the outflow. In fact, on top of this eclipse event, we also find stable absorption features present in both the eclipse and non-eclipse spectra (see Fig. \ref{fig:4}). This suggests that outflowing material is persistently present along the line of sight, but clumps within the flow occasionally enter the line of sight, causing additional absorption over a specific period, which manifests as an eclipse event.

In the future, multi-wavelength spectroscopic monitoring \citep[similar to the AGN STORM project, ][]{2015MNRAS.453..214D} can be designed to examine the shape of optical/UV absorption lines during eclipse events. Additionally, XRISM observations provide new insights into the origin of such eclipse events \citep{2024ApJ...973L..25X}. By studying the profile of the Iron K$\alpha$ line during eclipses, it may be possible to verify their link to the BLR. These observations could ultimately determine the potentially diverse origins of different eclipse events.

\subsection{Comparison with Previous Works}
\subsubsection{Individual Sources}

In this work, we identify 3 AGN eclipse events with high confidence and 8 candidates in the \textit{Swift}/XRT archival data. We compile 33 previously reported obscuration events within our AGN sample, listed in Table \ref{tab:appA}, along with their identification status by our method for comparison. These include eclipse events and some absorbers originated from outflows. However, they do not cover all outflow or warm absorber cases, as their timescales are either too short (less than one day) or too long (several months to years) for {\it Swift} to capture the ingress and egress phases. Among the 33 known obscuration events, 28 were not captured by {\it Swift}: five are not identified by our selection criteria, while the remaining 23 occurred during periods when {\it Swift} had no observations or the observations were not used in this work. The five missed events will be further explored in Section \ref{sec:5.3}.

Notably, the 5 eclipses identified by this work are well-connected to previous studies, mutually corroborating and supporting each other. Around the time of the eclipse event 1H 0707-495/2010.11, deep X-ray observations were conducted by \textit{XMM-Newton} and \textit{Chandra} in September and October 2010, as well as November 2011. Spectral analyses of these observations suggested the presence of highly ionized outflows \citep{2012MNRAS.422.1914D,2016MNRAS.460.1716D}. Similarly, \textit{XMM-Newton} performed a deep observation of NGC 6814 in April 2016, revealing a quick occultation event \citep{2021ApJ...908L..33G}, which occurred within the eclipse period of our candidate event NGC 6814/2016.4. The two candidates found in NGC 3227 were also reported and further investigated in previous studies \citep{2015A&A...584A..82B,2022A&A...665A..72M,2023A&A...673A..26G}. Our derived $N_{\rm H}$ of the candidate NGC 3227/2008.10 event is consistent with results in \citet{2015A&A...584A..82B}. The candidate identified in MR 2251-178 has been comprehensively analyzed by \citet{2022ApJ...940...41M}. 

We identify 6 eclipse events that were not reported in previous studies. For example, in Mrk 817, we report two new events, Mrk 817/2018.5 and the candidate Mrk 817/2018.3, both of which occurred in 2018. Subsequently, an obscuration event with a long duration (exceeding one year) and variable $N_{\text H}$ was detected at the end of 2020 \citep{2021ApJ...922..151K,2023ApJ...947....2P}. Regarding the candidate event Mrk 841/2013.12.28, the hardness ratio exceeded the 3$\sigma$ threshold, indicating strong soft X-ray absorption in the spectrum.

\subsubsection{Comparison with the RXTE sample}
\citetalias{2014MNRAS.439.1403M} conducted the first systematic search for AGN eclipse events in the X-ray band using historical {\it RXTE} observational data. They identified 12 events in 8 AGNs. We can make a detailed comparison with this sample study.

Our sample consists of 40 AGNs, including 33 type I AGNs (including both Seyfert 1 and 1.5 galaxies) and 7 type II AGNs (including both Seyfert 1.8 and 2 galaxies). We identify 11 eclipse events (including candidates) in 8 type I AGNs but find none in type II AGNs. In comparison, the \citetalias{2014MNRAS.439.1403M} sample includes 56 AGNs, with 37 type I and 18 type II AGNs. They identified 8 events in 5 type I AGNs and 4 events in 3 type II AGNs. The detection rate of eclipse events in type I AGNs is broadly consistent between the two sample studies. Given that our sample contains only 7 type II AGNs, our non-detection of eclipse events in these sources may be due to random fluctuations in the sample. Since the nature of eclipse events in type I and type II AGNs could differ significantly, and no new events were found in our type II AGNs, we focus primarily on comparing our results with the type I AGN subsample from the \citetalias{2014MNRAS.439.1403M} study in the following analysis.

First, compared to \citetalias{2014MNRAS.439.1403M}’s study based on the {\it RXTE} observations, our work using {\it Swift}/XRT benefits from a broader energy range reaching down to 0.3 keV, whereas {\it RXTE} is limited to hard X-rays above 2 keV. This makes our data more sensitive to the presence of transient obscurers.

Then, we compare the sampling of the light curves of the two samples. Both the {\it Swift} and {\it RXTE} monitoring of AGNs are non-uniform, and the sampling duration and cadence directly affect the timescales of the identified eclipse events. We adopt the method of 'selection function' as described in \citetalias{2014MNRAS.439.1403M}. This method involves selecting different timescales over a given time range and then examining the number of valid light-curve segments for each timescale. We adopt a series of 21 timescale bins ranging from 0.12 to 10030 days, evenly distribute in logarithmic space. In each timescale bin, a valid light-curve segment requires at least 4 data points, and the time interval between any two adjacent points must not exceed 75\% of the timescale. These two criteria are intended to ensure that each segment has sufficient and relatively uniform sampling. Then, for each timescale bin, we calculate the number of valid segments contained in our entire 33 type I AGNs, from which we obtain the summed selection function (SF$_{\text{sumI}}$), i.e., the count of valid segments in each timescale bin.

Additionally, since the light curves within each AGN may have significantly different sampling, we adopt the following method to quantitatively assess the degree of this difference:

{(1) Step 1:} We randomly select one source from the 33 type I AGNs of this study and repeat this selection 33 times to create a new sample containing 33 AGNs. Each AGN in this new sample has its real {\it Swift}/XRT light curve, but some AGNs may be repeated since they are all drawn from the same AGN sample. Then, we calculate the SF$_{\text{sumI}}$ for this new sample.

{(2) Step 2:} We repeat Step 1 for a total of 1000 times to obtain 1000 SF$_{\text{sumI}}$ values, from which we compute the standard deviation for each timescale bin.

The meaning of this method is that if the light curve sampling of every AGN in the sample were identical, then the 1000 SF$_{\text{sumI}}$ values should all be the same, and the standard deviation should be 0. Conversely, if the standard deviation is large, it indicates that there are significant sampling differences between individual AGNs in the sample. 

\begin{figure}[ht!]
\centering
\includegraphics[width=0.45\textwidth]{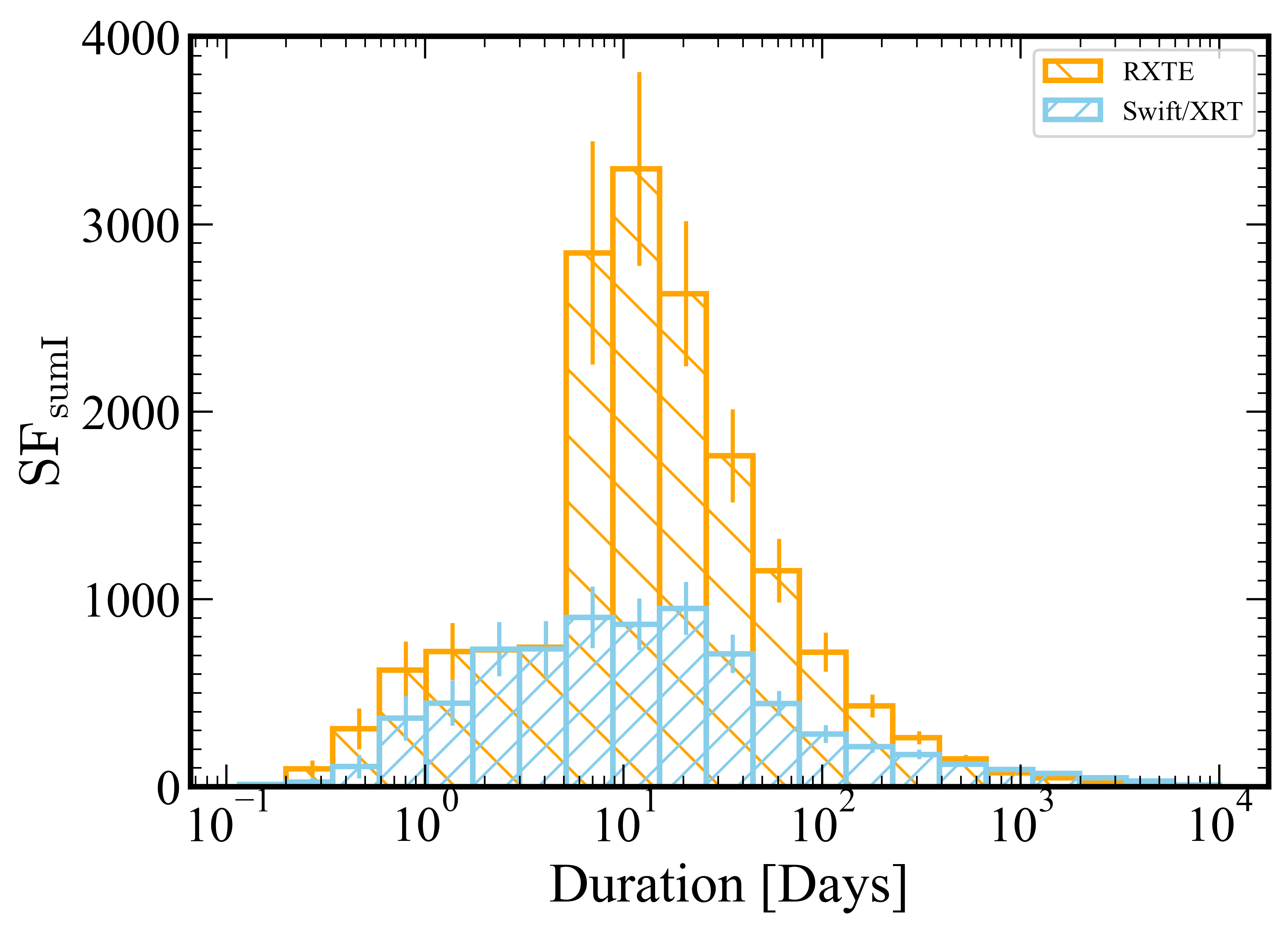}
\caption{The selection function SF$_{\text{sumI}}$ for 33 type I AGN of {\it Swift}/XRT and 37 type I AGN of \textit{RXTE} (\citetalias{2014MNRAS.439.1403M}). 
\label{fig:fun}}
\end{figure}

Fig. \ref{fig:fun} compares the SF$_{\text{sumI}}$ of our work and \citetalias{2014MNRAS.439.1403M}. It can be seen that the time span covered by our work is similar to \citetalias{2014MNRAS.439.1403M}, and both samples have more sampling in the 10-100 days timescale range. Indeed, most of the eclipse events found in both studies are also within this timescale range.

Then, we can roughly estimate the probability of detecting eclipse events (i.e., obscuring clouds) in each timescale bin. Firstly, we calculate the number of detected eclipse events in each duration $N_{\text{ecl}}$(t$_{\text i}$). The best-estimated value $N_{\text{ecl}}$(t$_{\text i}$) denoted by blue bars only considers the number of high-confidence events in their average timescales between the upper and lower limits, and this range is influenced by differences in sampling. In the calculation for minimum $N_{\text{ecl}}$(t$_{\text i}$), we consider high-confidence events in duration with the maximum value of SF$_{\text{sumI}}$(t$_{\text i}$). As for the maximum $N_{\text{ecl}}$(t$_{\text i}$), 5 events having well-constrained timescales are chosen to calculate the number in duration with the minimum value of SF$_{\text{sumI}}$(t$_{\text i}$). The distributions of best, minimum and maximum $N_{\text{ecl}}$(t$_{\text i}$) are presented in Fig. \ref{fig:fun1}. 

\begin{figure}[ht!]
\centering
\includegraphics[width=0.45\textwidth]{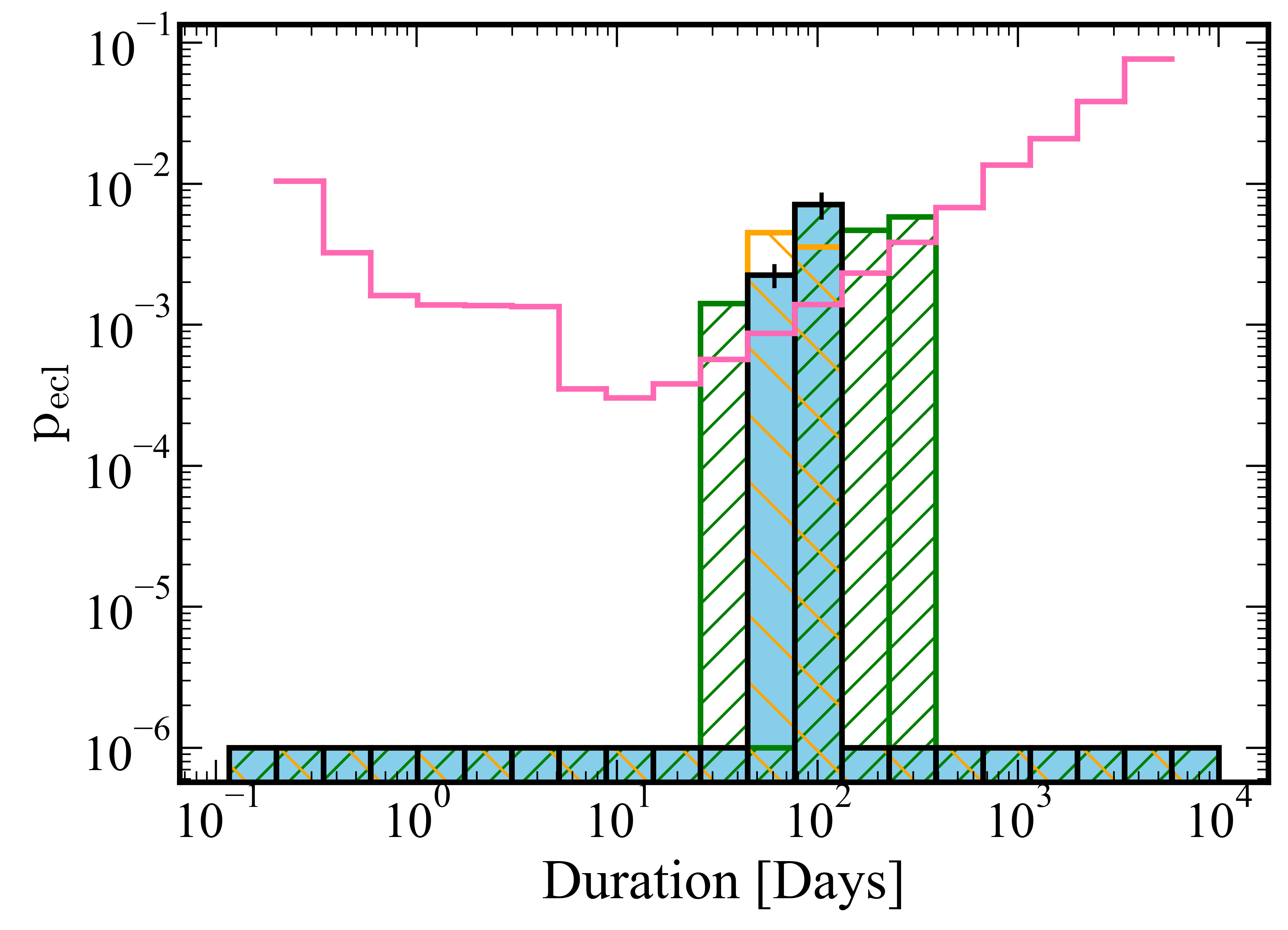}
\caption{The probability density function $p_{\text{ecl}}$(t$_{\text i}$) for the type I AGNs. The blue bar denotes the probability density derived by the best estimate $N_{\text{ecl}}$(t$_{\text i}$) with the error bar calculated by the uncertainty of SF$_{\text{sumI}}$(t$_{\text i}$). The orange and green solid lines represent the minimum and maximum probability densities, respectively. For durations with no eclipse event detection, we set the probabilities as 10$^{-6}$ to denote the zero value. The hot pink line denotes 1/SF$_{\rm sumI}$(t$_{\text i}$) from \textit{RXTE} (\citetalias{2014MNRAS.439.1403M}).
\label{fig:fun2}}
\end{figure}

Secondly, the obtained values of $N_{\text{ecl}}$(t$_{\text i}$) are divided by SF$_{\text{sumI}}$(t$_{\text i}$) in the same duration (t$_{\text i}$) to estimate the probability density function $p_{\text{ecl}}$(t$_{\text i}$). For comparison, we also present the \textit{RXTE} results of 1/SF$_{\rm sumI}$(t$_{\text i}$) from \citetalias{2014MNRAS.439.1403M}. As shown in Fig. \ref{fig:fun2}, the probability of observing an eclipse event on the 10--1000 day timescale is 0.001--0.01. However, due to the small number of detected eclipse events, the uncertainties in this small-sample statistic should be significant. Additionally, since our selection criteria for eclipse events prioritize high reliability over completeness, the observed probability must be considered a lower limit. Therefore, a more reliable conclusion may be that the observing probability for these medium-timescale eclipse events should be $> 0.001$, which is also broadly consistent with \citetalias{2014MNRAS.439.1403M}. 

\subsection{Limitations of the Searching Strategy and Results}
\label{sec:5.3}

Our search strategy favors identifying eclipses with $N_{\rm H}$ in the range of 10$^{21 - 23}$ cm$^{-2}$ and timescales from a few to hundreds of days, associated with neutral or mildly ionized clouds. However, due to the non-uniformly sampled {\it Swift}/XRT light curves and our selection criteria, this approach inevitably misses some obscuration events that have short-lived (less than a day) and long-lived (spanning months to years) timescales, as well as Compton-thick or highly ionized eclipsing clouds due to low signal-to-noise ratio and spectral resolution observations.

To begin with, as shown in Fig. \ref{fig:fun}, the light curves in our sample are primarily sampling in the 1--1000 day timescale range, and unsurprisingly, the timescales of the identified eclipse events also fall within this range. Therefore, regarding the timescale range of eclipse events, the results of this work are not complete. Additionally, due to the impact of non-uniform sampling, we lack effective constraints on the upper and lower limits of the duration for several eclipse events, making it difficult to provide a reliable estimate of observing probability.

Moreover, our potential eclipse selection primarily relies on hardness ratio deviations between the 0.3–2 and 2–10 keV bands, making it less sensitive to Compton-thick eclipses. These absorbers extend attenuation to harder bands and make the absorption pattern of our defined hard band similar to that in the soft band, reducing the amplitude of hardness ratio variations. Additionally, Compton-thick clouds can cause extreme flux drops, sometimes rendering sources undetectable by {\it Swift}/XRT and being excluded during the identification process, potentially leading to missed Compton-thick clouds when relying on light curves. Hard X-ray telescopes (e.g., {\it NuSTAR}) are better suited for detecting such clouds. Lastly, limiting spectral fitting to 6 keV may result in an underestimation of the derived $N_{\rm H}$ for eclipsing clouds.

Notably, some highly ionized clouds might be omitted by our hardness ratio selection. For example, in mid-2012, NGC 1365 experienced highly ionized absorption \citep{2014ApJ...788...76W}, where the intrinsic spectral shape varied with strong absorption near 2 keV. However, its soft X-ray emission ($<$ 1 keV) remained unaffected and appeared similar to faint absorbed periods, resulting in a stable hardness ratio value. Combined with insufficient {\it Swift} monitoring during this period, this event was not identified by our criteria.

Another selection bias arises from the use of a 2$\sigma$ deviation threshold in hardness ratio. Since we cannot predict in advance which light curve segment contains an eclipse, in our identification strategy we use the mean and standard deviation of the flux and hardness ratio of the entire light curve as selection criteria. However, the average hardness ratio may not adequately represent normal continuum variation because the occurrence of an eclipse event can significantly alter the flux and hardness ratio, increasing their standard deviation. As a result, using a $2\sigma$ threshold for selection may miss some relatively weaker eclipse events \citep[e.g. PDS 456/2017-3 ][]{2018ApJ...867...38R}.

Given the limitation of detection sampling and hardness ratio-based selection criteria, it is expected that no eclipses were identified in the seven Seyfert 2 galaxies within our 40 AGN sample. Additionally, our method does not capture long-timescale obscuration events. Some sources exhibit significant and variable absorption over extended periods while maintaining a relatively stable hardness ratio, often associated with outflows \citep[e.g. NGC 3783, NGC 5548 and Mrk 335 ][]{2017A&A...607A..28M,2014Sci...345...64K,2020A&A...643L...7K}. However, these cases do not match the characteristics of transient eclipses caused by clumpy material that we aim to identify. Therefore, we do not consider them as eclipse events. 

\section{Summary} \label{sec:summary}
In this work, we conduct a systematic search for AGN eclipse events based on approximately 11,000 observations of 40 AGNs (including 33 type I AGNs and 7 type II AGNs) monitored by {\it Swift}/XRT over the past 20 years. Our selection criteria require significant changes in both flux and hardness ratio, as well as significant spectral shape variations between the eclipse and non-eclipse phases, which could be effectively modeled by PCA. Our main results are summarized as follows:

1. We identify 3 high-confidence eclipse events in 3 AGNs and 8 additional candidates in 6 AGNs, all of which are type I AGNs. The timescales of these events range from a few days to over a year, with column densities spanning  $(0.2 - 31.2) \times 10^{22}$ \, \text{cm}$^{-2}$  and covering factors ranging from 0.6 to 1. The absorbers observed in our sample are predominantly neutral or modestly ionized, with $\log \xi$ values ranging from -1.3 to 2.2.

2. Based on the assumption of Keplerian orbits and spectral fitting results, we constrain the distances $r_{\text{c}}$  of the obscuring clouds from the central black hole to be $(2.4-179) \times 10^4~R_{\rm g}$, and the cloud diameter ranges from 0.05 to 1 light-days for the 5 eclipse events with well-constrained duration. We also calculate the outer radius $R_{\text{d}}$ of the DSZ and compared it with the cloud distance, yielding $r_{\text{c}} / R_{\text{d}}$ ratios varying between 0.08 and 11.51. Among the well-constrained duration events, one of them is closer to the nucleus than the DSZ, making it potentially dust-free, and indeed more ionized than the other events. The rest are located inside and outside the DSZ, exhibiting a lower ionization state.

3. We find that the physical parameters of these eclipse events, including column density and ionization state, exhibit certain correlations with the $M_{\rm BH}$. This suggests that the formation of these clouds is directly linked to the black hole accretion process, probably associated with clumpy outflows.

4. Due to the incompleteness of our eclipse event sample, we could only constrain the observing probability of eclipse events on the 10--1000 day timescale to be $>$ 0.001, which is broadly consistent with the results of previous works.

The results of this work complement the existing sample of AGN eclipse events and contribute to the studies of AGN outflows and the environment around their nuclei. With more observations coming from the newly launched wide-field X-ray all-sky monitor, {\it EP} \citep{Yuan2022}, we will be able to monitor hundreds of nearby AGNs over years in the 0.5--4 keV band \citep{2025arXiv250107362Y}. This will enable a more systematic determination of the occurrence rates of these eclipse events and a deeper understanding of their connection to AGN accretion processes.\\

We thank the anonymous referee for providing valuable comments and suggestions to improve this work. We thank Johnathon Gelbord for the valuable discussion about the {\it Swift}/UVOT data analysis.  C.J. acknowledges the National Natural Science Foundation of China through grant 12473016, and the support by the Strategic Priority Research Program of the Chinese Academy of Sciences (Grant No. XDB0550200). This work is based on observations conducted by XRT and UVOT on board the Neil Gehrels Swift Observatory, the data is provided by the High Energy Astrophysics Science Archive Research Center (HEASARC), a service of the Astrophysics Science Division at NASA/GSFC. 



\bibliography{sample631}{}
\bibliographystyle{aasjournal}

\appendix

\section{The Sample}
\label{app:A}

Our original AGN sample is selected based on the catalogue from \citet{2022ApJ...936..105H} and \textit{RXTE} AGN Timing \& Spectral Database \citep{2013ApJ...772..114R}. We only include sources that have been detected by XRT for more than 90 observations or continuous monitoring for at least 50 observations. The detection significance of sources should be greater than 3$\sigma$ to calculate the hardness ratio and perform further study of the eclipse event. This table also summarizes the historical obscuration events from the previous studies.

\newcommand{\Myref}{\citet{2024A&A...684A.167G}}
\newcommand{\MyrefM}{\citet{2017A&A...607A..28M}}
\newcommand{\MyrefK}{\citet{2020A&A...643L...7K}}
\newcommand{\MyrefT}{\citet{2008A&A...483..161T}}
\newcommand{\MyrefG}{\citet{2023A&A...673A..26G}}
\newcommand{\MyrefL}{\citet{2010A&A...510A..92L}}
\newcommand{\MyrefMo}{\citet{2010A&A...517A..47M}}
\newcommand{\MyrefP}{\citet{2010A&A...510A..65P}}
\newcommand{\MyrefB}{\citet{2015A&A...584A..82B}}
\newcommand{\MyrefW}{\citet{2022A&A...657A..77W}}

\startlongtable
\begin{deluxetable*}{CCCCCCCCC} 
\tablenum{A}
\setlength{\tabcolsep}{1pt}
\tablecaption{Table of the 40 sources in our AGN sample
\label{tab:appA}}
\tablehead{
\colhead{Source Name} & \colhead{Type}& \colhead{Redshift} & \colhead{log$M_{\rm BH}$} & \colhead{Ref\,$^a$} & \colhead{Exp\,$^b$} & \colhead{ObsIDs\,$^c$} & \colhead{Known Obscuration Events\,$^d$} & \colhead{Identification\,$^e$} \\
\colhead{} & \colhead{} & \colhead{} & \colhead{($M_{\odot}$)} & \colhead{} & \colhead{(ks)} & \colhead{} & \colhead{} & \colhead{(\textnormal{Y/N})}
}
\startdata
\textnormal{3C 120}	&	\textnormal{Sy1/BLRG}	&	0.033 	&	7.74$^{\ddag}$ 	&	1	&	332	&	237	&		\\
\textnormal{3C 273}	&	\textnormal{Sy1/FSRQ}	&	0.158 	&	8.61$^{\ddag}$ 	&	2	&	626	&	305	&		\\
\textnormal{Ark 120}	&	\textnormal{Sy1}	&	0.033 	&	8.18$^{\ddag}$ 	&	1	&	149	&	138	&		\\
\textnormal{Fairall 9}	&	\textnormal{Sy1}	&	0.047 	&	8.41$^{\ddag}$ 	&	1	&	639	&	809	&  	\textnormal{Eclipse 2001-3 (M14)} & \textnormal{N(i)} \\
\textnormal{MCG+8-11-11}	&	\textnormal{Sy1}	&	0.021 	&	7.29$^{\ddag}$ 	&	3	&	379	&	386	&		\\
\textnormal{Mrk 279}	&	\textnormal{Sy1}	&	0.030 	&	7.54$^{\ddag}$ 	&	1	&	494	&	381	&		\\
\textnormal{NGC 3783}	&	\textnormal{Sy1}	&	0.010 	&	7.47$^{\ddag}$ 	&	1	&	212	&	186	&	\textnormal{Eclipse 2008-3 (M14)} & \textnormal{N(i)}\\
& & & & & & & \textnormal{Eclipsing clumpy wind in 2016-12 (M17)} & \textnormal{N(ii)}\\
\textnormal{NGC 4593}	&	\textnormal{Sy1}	&	0.009 	&	6.99$^{\ddag}$ 	&	1	&	268	&	279	&	\textnormal{Warm absorber (E13)}	& \textnormal{N(i)}\\
\textnormal{PDS 456} &	\textnormal{Sy1}	&	0.184 	&	8.23$^{\dag}$ 	&	4	&	507	&	185	&	\textnormal{Eclipse 2017-3 (R18)} & \textnormal{N(ii)}	\\
\textnormal{1H 0707-495}	&	\textnormal{NLS1}	&	0.041 	&	6.31$^{*}$ 	&	5	&	180	&	162	&	\textnormal{Highly ionized outflow (D12/C16)} & \textnormal{Y}\\
\textnormal{Ark 564}	&	\textnormal{NLS1}	&	0.025 	&	6.06$^{*}$	&	5	&	100	&	89	&		\\
\textnormal{IRAS 13224-3809}	&	\textnormal{NLS1}	&	0.066 	&	6.82$^{*}$ 	&	5	&	134	&	105	&		\\
\textnormal{Mrk 110}	&	\textnormal{NLS1}	&	0.035 	&	7.40$^{\ddag}$ 	&	1	&	504	&	466	&		\\
\textnormal{Mrk 335}	&	\textnormal{NLS1}	&	0.026 	&	7.15$^{\ddag}$ 	&	1	&	1118	&	1024	&	\textnormal{Partial covering absorber 2020-5 (K20)} & \textnormal{N(ii)}	\\
\textnormal{Mrk 766}	&	\textnormal{NLS1}	&	0.013 	&	6.82$^{\ddag}$ 	&	3	&	149	&	84	&	\textnormal{Eclipse 2005-5 (R11a)} & \textnormal{N(i)}\\
\textnormal{PKS 0558-504}	&	\textnormal{NLS1}	&	0.137 	&	7.78$^{*}$ 	&	6	&	180	&	90	&		\\
\textnormal{RE J1034+396}	&	\textnormal{NLS1}	&	0.042 	&	6.81$^{*}$ 	&	5	&	102	&	108	&		\\
\textnormal{RX J0134.2-4258}	&	\textnormal{NLS1}	&	0.237 	&	7.30 	&	7	&	165	&	101	&		\\
\textnormal{ESO 323-G77}	&	\textnormal{Sy1.2}	&	0.015 	&	7.39$^{\star}$ 	&	8	&	139	&	100	&		\\
\textnormal{Mrk 590}	&	\textnormal{Sy1.2}	&	0.026 	&	7.68$^{\ddag}$ 	&	1	&	555	&	263	&		\\
\textnormal{NGC 7469}	&	\textnormal{Sy1.2}	&	0.016 	&	7.09$^{\ddag}$ 	&	1	&	519	&	641	&		\\
\textnormal{MCG-6-30-15}	&	\textnormal{NLS1.2}	&	0.008 	&	6.30$^{\ddag}$ 	&	3	&	433	&	428	&		\\
\textnormal{MR 2251-178}	&	\textnormal{Sy1.5}	&	0.064 	&	8.28$^{\star}$ 	&	9	&	91	&	95	&	\textnormal{Eclipse 1996 (M14)} &	\textnormal{N(i)} \\ & & & & & & & \textnormal{and late-2020 to early-2021 (M22)} & \textnormal{Y} \\
\textnormal{Mrk 509}	&	\textnormal{Sy1.5}	&	0.034 	&	8.00$^{\dag}$ 	&	4	&	330	&	325	&	\textnormal{Eclipse 2005-9 (M14)} & \textnormal{N(i)}	\\  
\textnormal{Mrk 817}	&	\textnormal{Sy1.5}	&	0.031 	&	7.59$^{\ddag}$ 	&	3	&	185	&	194	&	\textnormal{Ionized obscurer in 2020-2022 (K21)}	& \textnormal{N(iii)} \\
\textnormal{Mrk 841}	&	\textnormal{Sy1.5}	&	0.036 	&	8.15$^{*}$ 	&	5	&	216	&	136	&	\textnormal{Warm absorber in 2001 and 2005 (L10)} & \textnormal{N(i)}\\
\textnormal{NGC 1566}	&	\textnormal{Sy1.5}	&	0.005 	&	6.11$^{\star}$ 	&	8	&	319	&	263	&		\\
\textnormal{NGC 3227}	&	\textnormal{Sy1.5}	&	0.004 	&	7.63$^{\ddag}$ 	&	1	&	137	&	92	&	\textnormal{Obscuration event  2000-1 (L03)} & \textnormal{N(i)}\\ & & & & & & & \textnormal{2002-8 (M14)} & \textnormal{N(i)} \\ & & & & & & & \textnormal{2006-12 (W22)} & \textnormal{N(i)} \\ & & & & & & & \textnormal{2008 (B15)} & \textnormal{Y} \\ & & & & & & & \textnormal{2016-12 (T18)} & \textnormal{N(i)} \\ & & & & & & & \textnormal{and end of 2019 (G23)} & \textnormal{Y}	\\
\textnormal{NGC 3516}	&	\textnormal{Sy1.5}	&	0.009 	&	7.40$^{\ddag}$ 	&	3	&	237	&	238	&	\textnormal{High column absorber in 2001 (T05)} & \textnormal{N(i)}\\ & & & & & & & \textnormal{Eclipse 2006-10 (T08)} & \textnormal{N(i)} \\ & & & & & & & \textnormal{and 2011-7 (M14)} & \textnormal{N(i)}\\
\textnormal{NGC 4151}	&	\textnormal{Sy1.5}	&	0.003 	&	7.12$^{\ddag}$ 	&	1	&	654	&	552	&	\textnormal{Obsorbers 1996 and 2001 (P07)}& \textnormal{N(i)}	\\ & & & & & & & \textnormal{and 2008-3 (W10)} & \textnormal{N(i)}\\
\textnormal{NGC 5548}	&	\textnormal{Sy1.5}	&	0.017 	&	7.70$^{\ddag}$ 	&	3	&	847	&	892	&	\textnormal{A outflow in 2013-2014 (K14)}	& \textnormal{N(ii)}\\
\textnormal{NGC 6814}	&	\textnormal{Sy1.5}	&	0.005 	&	7.04$^{\ddag}$ 	&	3	&	394	&	460	&	\textnormal{Eclipse 2016 (G21)} & \textnormal{Y}\\
\textnormal{NGC 4051}	&	\textnormal{NLSy1.5}	&	0.002 	&	6.28$^{\ddag}$ 	&	1	&	62	&	60	&		\\
\textnormal{NGC 1365}	&	\textnormal{Sy1.8}	&	0.006 	&	6.30$^{\star}$ 	&	10	&	343	&	169	&	\textnormal{Eclipse 2004-1 (R09)} & \textnormal{N(i)} \\ & & & & & & & \textnormal{ 2006-4 (R07)} &  \textnormal{N(i)}	\\ & & & & & & & \textnormal{and 2007-1 (M10)} &  \textnormal{N(i)}	\\ & & & & & & & \textnormal{Obscuration event 2012-2013 (W14)} &  \textnormal{N(ii)}	\\ 
\textnormal{NGC 4395}	&	\textnormal{Sy1.8}	&	0.001 	&	5.45$^{\ddag}$ 	&	3	&	381	&	304	& \textnormal{Eclipse 2003-11 (N11)} & \textnormal{N(i)}\\
\textnormal{IC 3599}	&	\textnormal{Sy1.9}	&	0.021 	&	6.85 	&	11	&	219	&	89	&		\\
\textnormal{1ES 1927+654}	&	\textnormal{Sy2}	&	0.017 	&	6.14$^{\star}$ 	&	12	&	143	&	125	&		\\
\textnormal{Cen A}	&	\textnormal{Sy2/blazar}	&	0.002 	&	7.74 	&	13	&	204	&	147	&	\textnormal{Eclipses 2003-2004 (M14)} & \textnormal{N(i)} \\ & & & & & & & \textnormal{and 2010-1 (R11b/M14)} & \textnormal{N(i)}\\
\textnormal{Circinus}	&	\textnormal{Sy2}	&	0.001 	&	6.23 	&	14	&	221	&	151	&		\\
\textnormal{NGC 2992}	&	\textnormal{Sy2}	&	0.008 	&	7.27$^{\star}$ 	&	15	&	237	&	130	&		\\
\enddata
\tablecomments{$^a$The black hole mass references:(1)\citet{2004ApJ...613..682P}; (2)\citet{2019ApJ...876...49Z}; (3)\citet{2015PASP..127...67B}; (4)\Myref; (5)\citet{2003MNRAS.343..164B}; (6)\MyrefP; (7)\citet{2023MNRAS.518.6065J}; (8)\citet{2007ApJ...660.1072W}; (9)\citet{2011MNRAS.415.1290L}; (10)\citet{2009ApJ...696..160R}; (11)\citet{2015ApJ...803L..28G}; (12)\citet{2022ApJ...933...70L}; (13)\citet{2009MNRAS.394..660C}; (14)\citet{2003ApJ...590..162G}; (15)\citet{2010MNRAS.402.1081V}. $^b$Total exposure time of XRT observation IDs whose detection significance {$>$} 3. $^c$The XRT observation IDs which the detection significance of the source {$>$} 3. $^d$The previously reported obscuration events, including eclipse/occultation events and some known absorbers that originated from outflows. However, we do not include all outflows or warm absorber events. The references are: M14=\citet{2014MNRAS.439.1403M}; M17=\MyrefM; E13=\citet{2013MNRAS.435.3028E}; R18=\citet{2018ApJ...867...38R}; K20=\MyrefK; R11a=\citet{2011MNRAS.410.1027R}; M22=\citet{2022ApJ...940...41M}; D12=\citet{2012MNRAS.422.1914D}; C16=\citet{2016MNRAS.460.1716D}; P07=\citet{2007MNRAS.377..607P}; W10=\citet{2010ApJ...714.1497W}; T05=\citet{2005ApJ...618..155T}; T08=\MyrefT; L03=\citet{2003MNRAS.342L..41L}; W22=\MyrefW; B15=\MyrefB; T18=\citet{2018MNRAS.481.2470T}; G23=\MyrefG; K14=\citet{2014Sci...345...64K}; L10=\MyrefL; K21=\citet{2021ApJ...922..151K}; G21=\citet{2021ApJ...908L..33G}; R09=\citet{2009ApJ...696..160R} ; R07=\citet{2007ApJ...659L.111R}; M10=\MyrefMo; W14=\citet{2014ApJ...788...76W}; N11=\citet{2011MNRAS.417.2571N}; R11b=\citet{2011ApJ...742L..29R}. $^e$Whether the event is identified by {\it Swift}. 'N' indicates that the event is not identified in this work due to : (i) the event occurring outside detection periods of {\it Swift}; (ii) it not being identified by our method; (iii) For Mrk 817, we only utilize the observations taken before the AGN-STORM 2 monitoring campaign. \\
$^{\dag}$ $M_{\rm BH}$ was measured through GRAVITY. \\
$^{\ddag}$ $M_{\rm BH}$ was measured through reverberation mapping. \\
$^{*}$ $M_{\rm BH}$ was measured as virial mass measured from the broad-line region.\\
$^{\star}$ $M_{\rm BH}$ was measured by the empirical correlation law. The rest of $M_{\rm BH}$ were estimated from other methods.
}
\end{deluxetable*}

\section{details of the candidate eclipse event in individual objects}
\label{app:B}

We show the X-ray flux and hardness ratio light curves for the identified candidate eclipse events in NGC 6814, NGC 3227, NGC 3783, Mrk 841 and MR2251-178. The candidate event Mrk 817/2018.3 is displayed in Fig. \ref{fig:1} and Fig. \ref{fig:2} as it is close to the high-confidence event Mrk 817/2018.5. To determine the candidate event, we also exhibit the comparison of spectral shape during the non-eclipse and eclipse phases. For some eclipses, due to insufficient sampling, we cannot estimate the upper and lower limit of eclipse duration; instead, we apply a grey-shaded region to display the observations chosen for stacked spectra. The best-fitting values of intrinsic parameters with 1$\sigma$ uncertainties are listed in Table \ref{tab:appB}, while the time-resolved spectral analysis results are shown in Table \ref{tab:appC}.

\begin{figure}[ht!]
\centering
\includegraphics[width=0.6\textwidth]{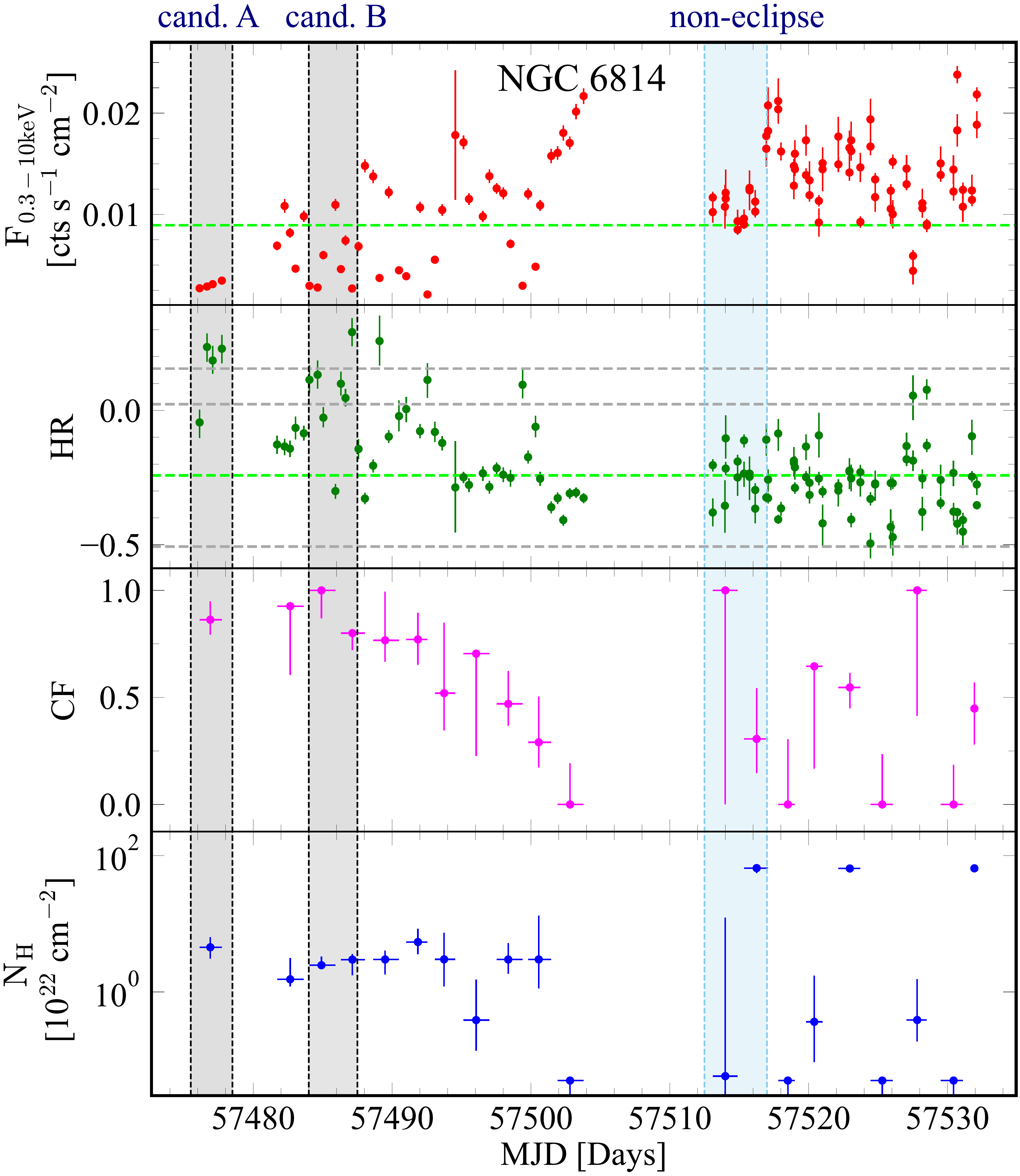}
\caption{The light curve of X-ray, hardness ratio, and time-resolved spectrum fitting parameters within 1$\sigma$ uncertainties for NGC 6814. The time-resolved spectral fitting is performed using a 2-day bin. The highlighted regions indicate the selected periods for stacking spectra of the non-eclipse phase (blue) and two candidate events (grey), NGC 6814/2016.3.30 and NGC 6814/2016.4.
\label{fig:NGC6814_lc}}
\end{figure}

\begin{figure}[ht!]
\centering
\includegraphics[width=0.45\textwidth]{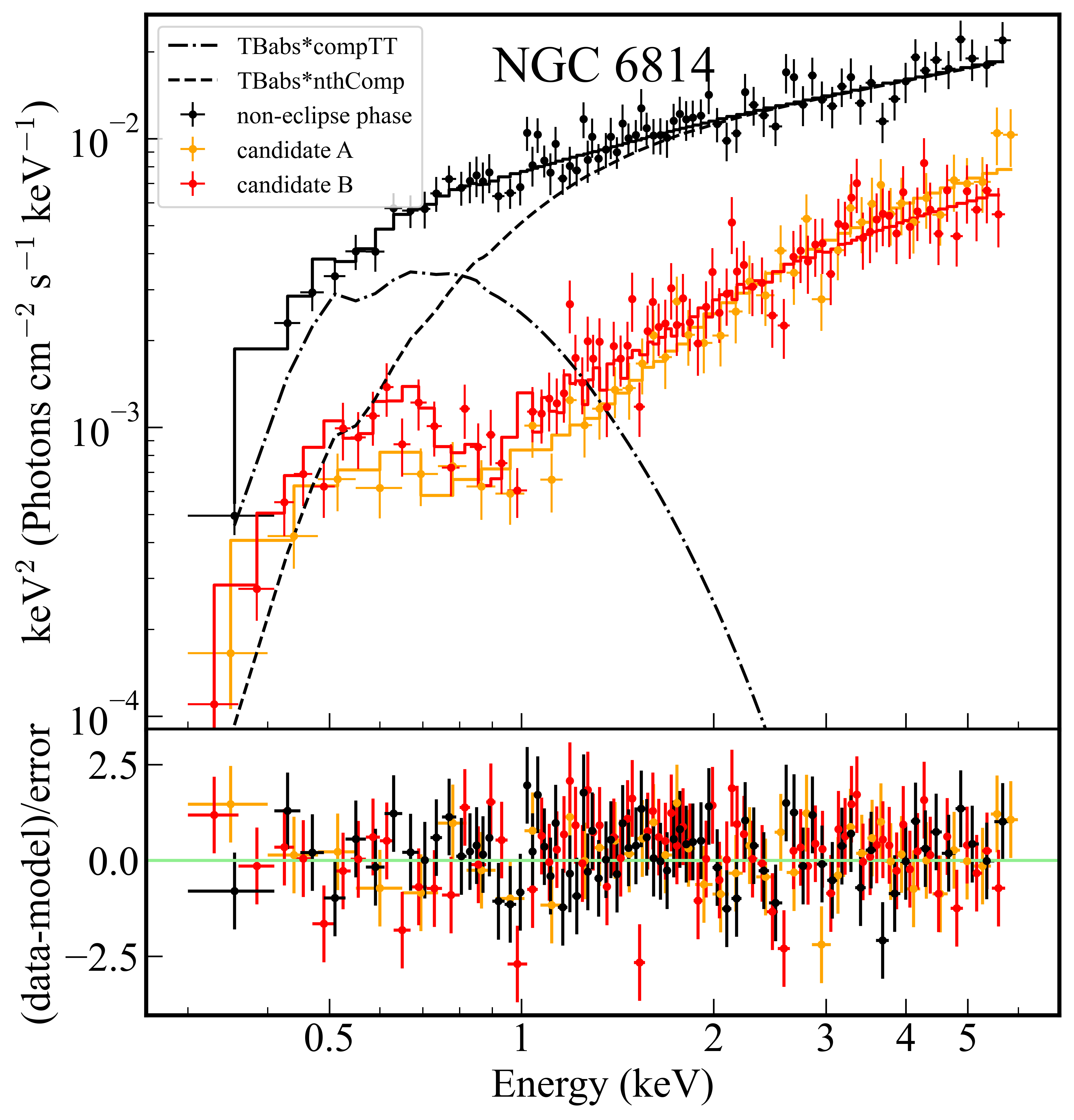}
\caption{The best fitting results of NGC 6814 during the non-eclipse and eclipse phases, the value of parameters are listed in Table \ref{tab:1} and Table \ref{tab:appB}.
\label{fig:NGC6814_spec}}
\end{figure}

\subsection{NGC 6814}

The light curves and spectra of two candidate events, NGC 6814/2016.3.30 and NGC 6814/2016.4, are depicted in Fig. \ref{fig:NGC6814_lc} and \ref{fig:NGC6814_spec}. Due to the high frequency of monitoring, we opt for a 2-day bin to carry out the time-resolved spectral analysis during these two eclipse intervals. The hardness ratios during eclipse periods are relatively high, accompanied by lower flux values compared to the non-eclipse phase, exceeding the 2$\sigma$ levels with some fluctuations. Moreover, the spectral shape during the eclipse event significantly differs from that of the non-eclipse phase, exhibiting strong absorption in the soft X-ray range, which can be well fitted by adding the PCA \texttt{zxipcf} model to the non-eclipse phase model (see Section \ref{sec:3.2} for details). 

During the eclipse-occurring period of candidate B, the time-resolved variations of $N_{\rm H}$ and CF significantly differ from those in the later, non-eclipse phase. Notably, an eclipse event during this interval was also independently reported by \citet{2021ApJ...908L..33G} based on \textit{XMM-Newton} observations.

\begin{figure}[ht!]
\centering
\includegraphics[width=0.6\textwidth]{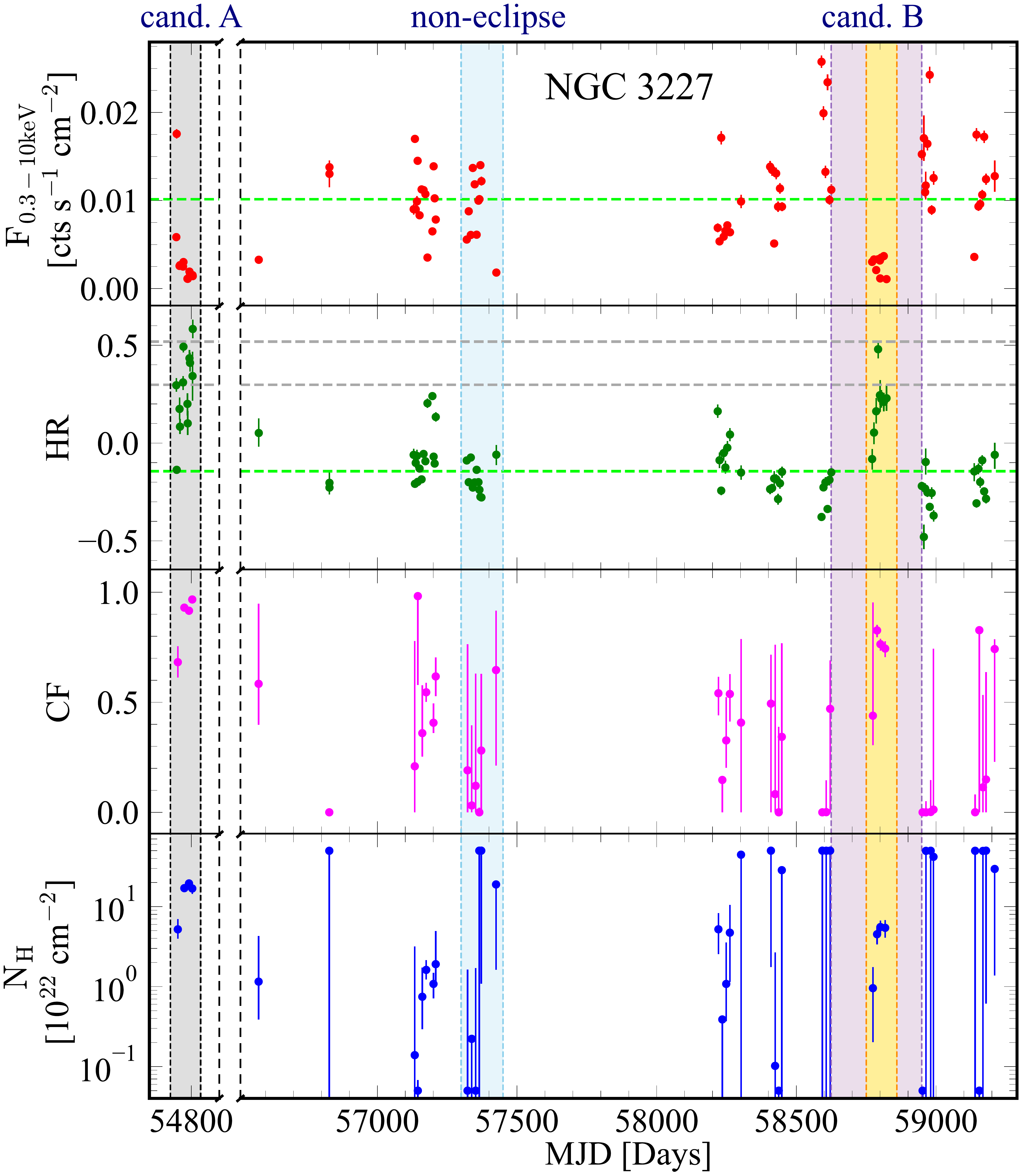}
\caption{The light curve of X-ray, hardness ratio, and time-resolved spectrum fitting parameters within 1$\sigma$ uncertainties for NGC 3227. The time-resolved spectral fitting is performed using a 10-day bin. The highlighted regions indicate the selected periods for stacking spectra of the non-eclipse phase (blue) and NGC 3227/2008.10 (cand. A, grey). The gold-shaded area represents the lower limit of the duration for NGC 3227/2019.10 (cand. B). The combination of this gold-shaded area and the purple-shaded areas denote the upper limit duration of the eclipse for NGC 3227/2019.10 (cand. B).
\label{fig:NGC3227_lc}}
\end{figure}

\begin{figure}[ht!]
\centering
\includegraphics[width=0.45\textwidth]{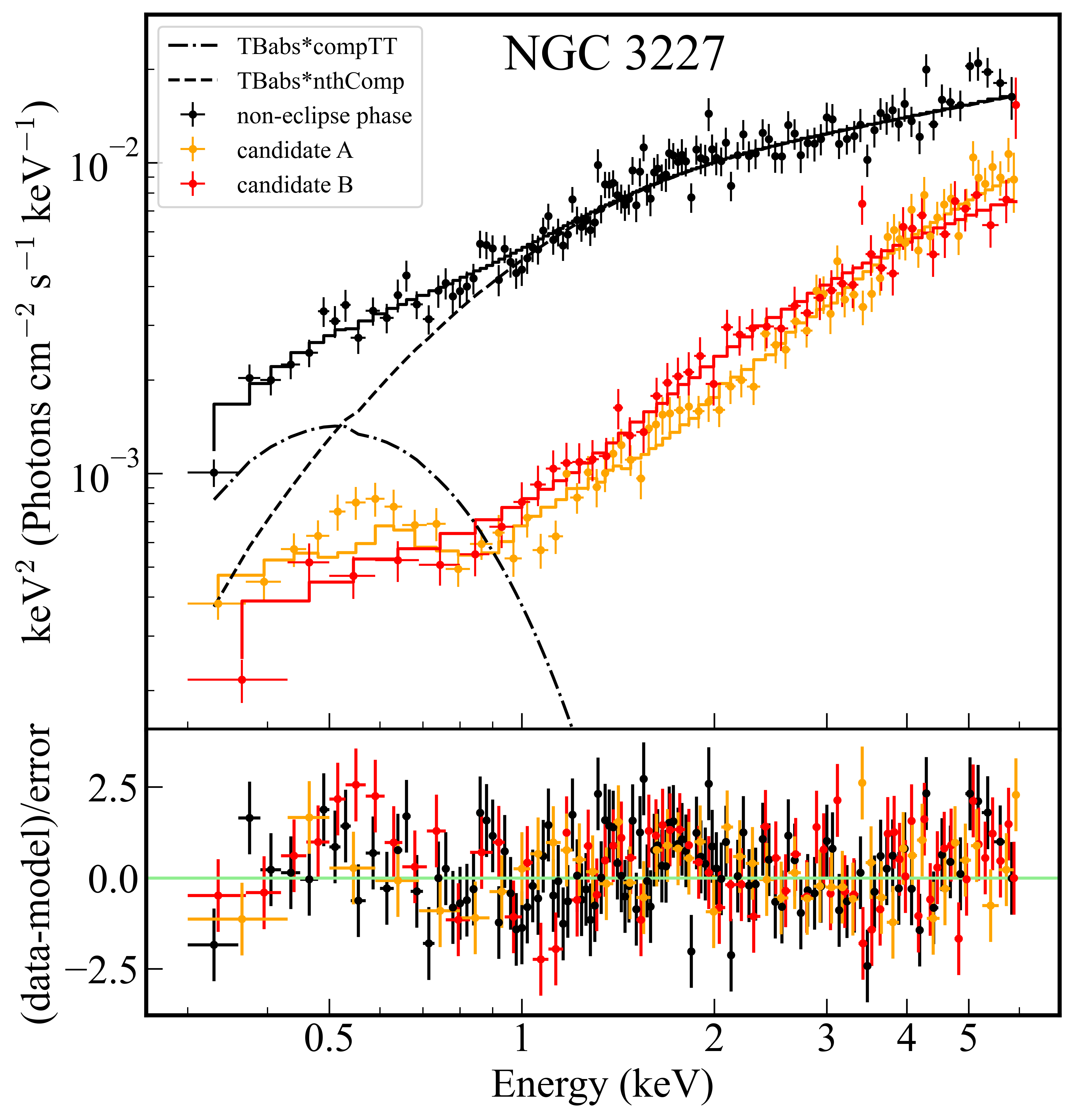}
\caption{The best fitting results of NGC 3227 during the non-eclipse and eclipse phases, the value of parameters are listed in Table \ref{tab:1} and Table \ref{tab:appB}.
\label{fig:NGC3227_spec}}
\end{figure}

\subsection{NGC 3227}

The light curves and spectra of two candidate events, NGC 3227/2008.10 and NGC 3227/2019.10, are depicted in Fig. \ref{fig:NGC3227_lc} and \ref{fig:NGC3227_spec}. The hardness ratios during the eclipse event are relatively high with lower flux values compared to the non-eclipse period. For candidate A, we select the entire continuous segment of the light curve, which is highlighted in grey, as the eclipse phase to improve the signal-to-noise ratio of the spectrum. In candidate B, we estimate the upper and lower limit of eclipse duration highlighted in light purple and gold, respectively. However, due to the non-uniform sampling, we choose a 10-day binning time to conduct the time-resolved spectrum fitting. These candidate events were also identified in previous studies, such as \citet{2015A&A...584A..82B, 2022A&A...665A..72M, 2023A&A...673A..26G}.

\subsection{NGC 3783}
\begin{figure}[ht!]
\centering
\includegraphics[width=0.6\textwidth]{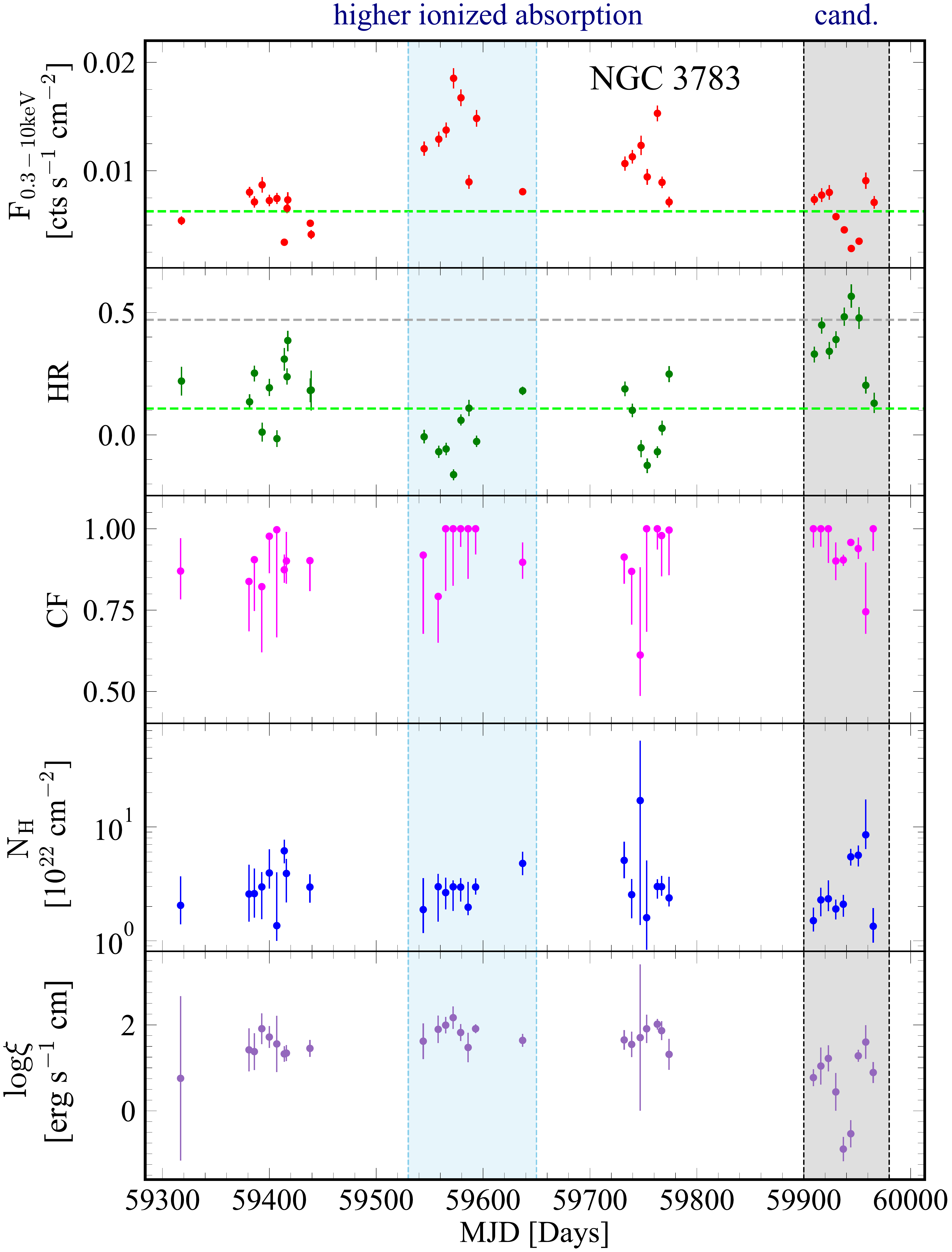}
\caption{The light curve of X-ray, hardness ratio, and time-resolved spectrum fitting parameters within 1$\sigma$ uncertainties for NGC 3783. A higher ionized absorption component is prominent in our defined non-eclipse spectrum (see Fig. \ref{fig:NGC3783_spec}). Therefore, we also analyze log $\xi$ in the time-resolved spectral fitting. This analysis is performed for each observation during both eclipse and non-eclipse periods. The highlighted regions indicate the selected periods for stacking spectra of non-eclipse phase (blue) and candidate event NGC 3783/2022.12 (grey).
\label{fig:NGC3783_lc}}
\end{figure}

\begin{figure}[ht!]
\centering
\includegraphics[width=0.45\textwidth]{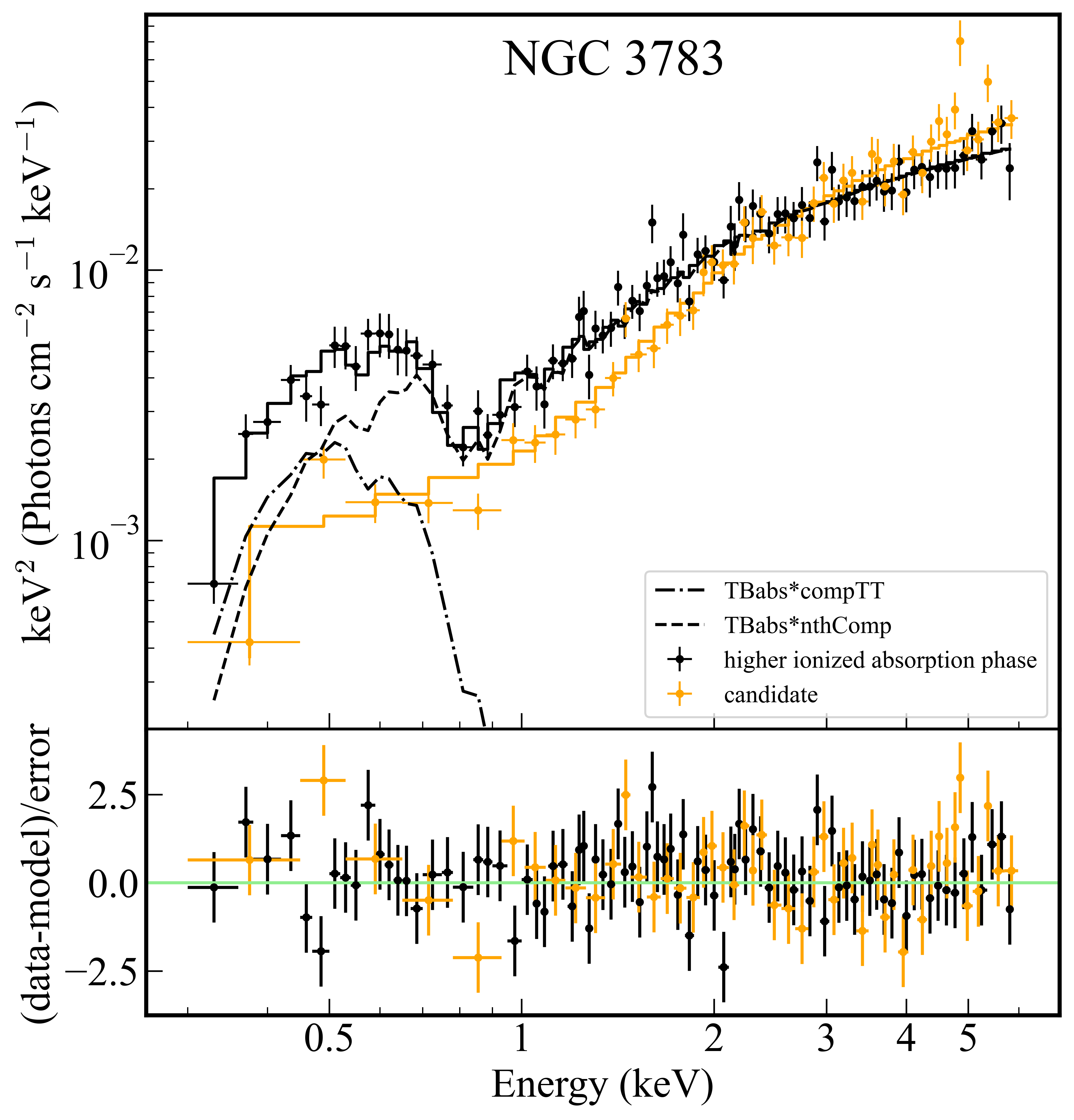}
\caption{The best fitting results of NGC 3783 during the high-ionized absorption state (defined non-eclipse phase) and eclipse phases, the value of parameters are listed in Table \ref{tab:1} and Table \ref{tab:appB}.
\label{fig:NGC3783_spec}}
\end{figure}

NGC 3783 is widely studied for its ionized flows and warm absorber \citep[e.g.][]{2002ApJ...574..643K,2006A&A...451L..23G,2017A&A...607A..28M}. An absorption component with higher ionization states is prominent in our defined non-eclipse phase spectrum (see Fig. \ref{fig:NGC3783_spec}). Thus, we let the log $\xi$, $N_{\rm H}$ and CF be free while analyzing the time-resolved spectra, the results are shown in Fig. \ref{fig:NGC3783_lc}. Since NGC 3783 is bright enough, to maintain the time resolution, we do not bin and stack the spectra during time-resolved spectral fitting. The hardness ratios during the candidate event exceed the 2$\sigma$ threshold, accompanied by a significant decrease in flux. We select the entire continuous observations of the light curve section (highlighted in grey) as the eclipse phase to improve the signal-to-noise ratio of the spectrum. Our fitting for the non-eclipse and candidate staking spectra are displayed in Fig. \ref{fig:NGC3783_spec}.

\subsection{Mrk 841}

\begin{figure}[ht!]
\centering
\includegraphics[width=0.6\textwidth]{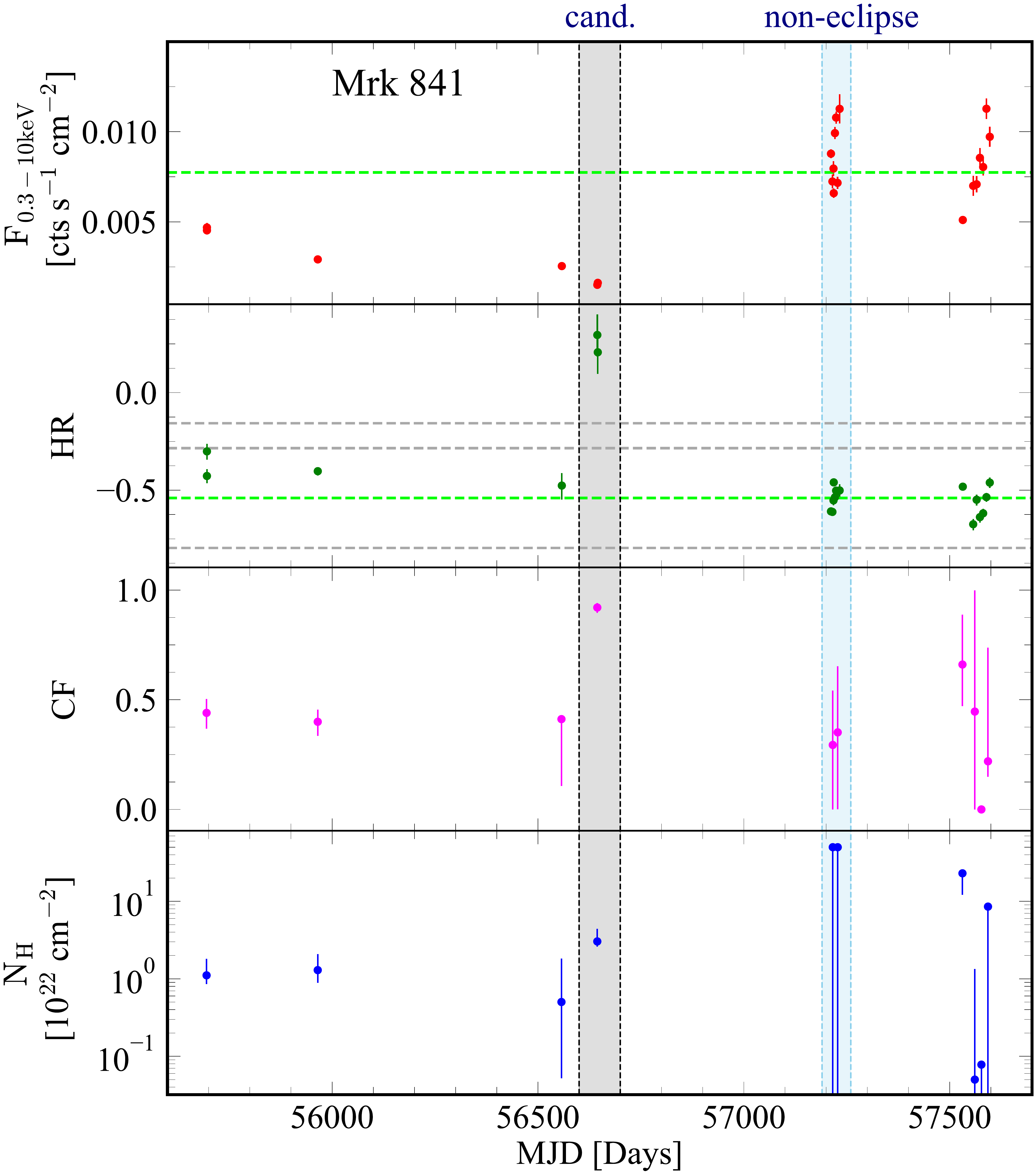}
\caption{The light curve of X-ray, hardness ratio, and time-resolved spectrum fitting parameters within 1$\sigma$ uncertainties for Mrk 841. The time-resolved spectral fitting is performed using a 10-day bin. The highlighted regions indicate the selected periods for stacking spectra of non-eclipse phase (blue) and candidate events Mrk 841/2013.12.28 (grey).
\label{fig:Mrk841_lc}}
\end{figure}

\begin{figure}[ht!]
\centering
\includegraphics[width=0.45\textwidth]{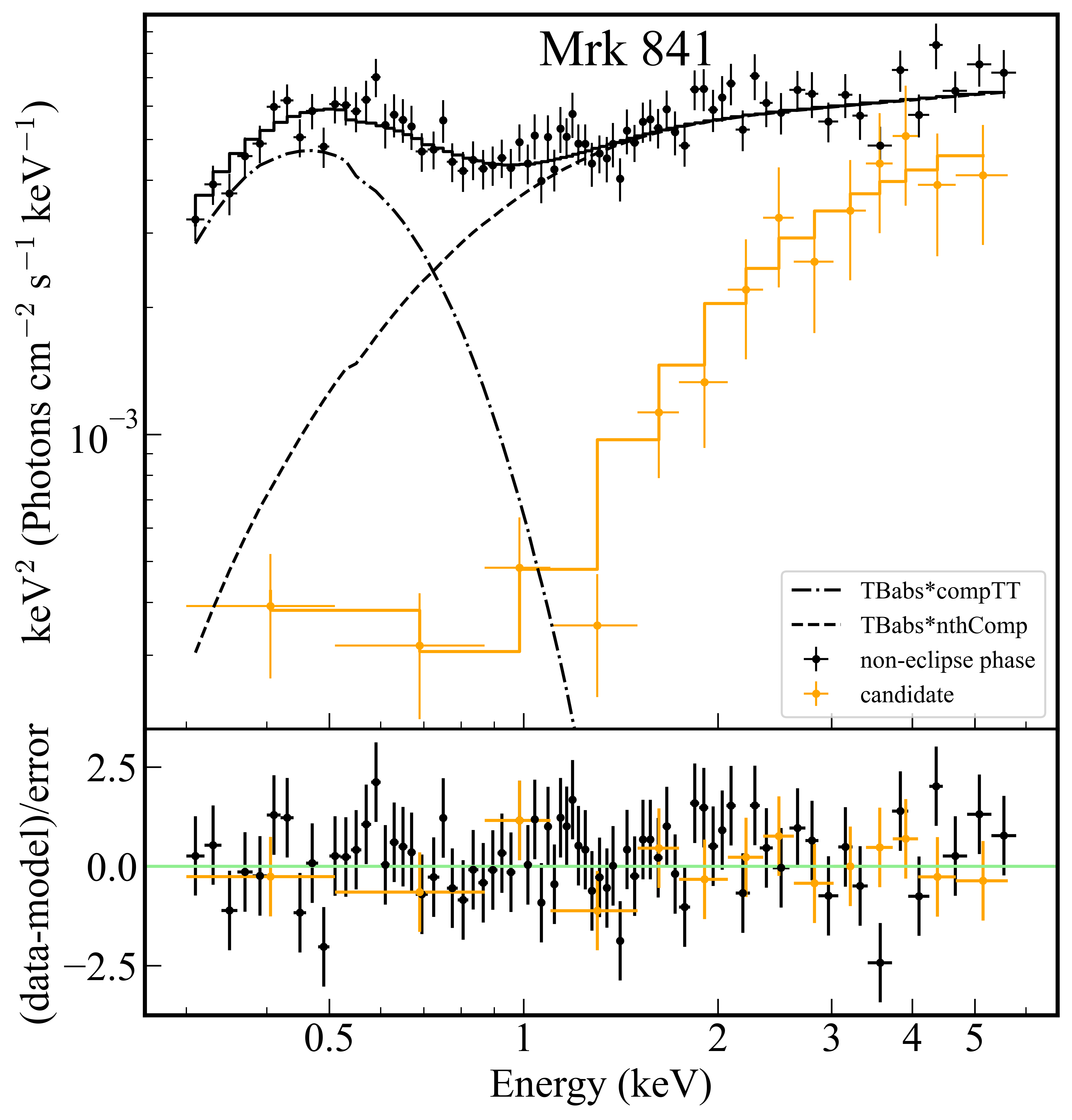}
\caption{The best fitting results of Mrk 841 during the non-eclipse and eclipse phases, the value of parameters are listed in Table \ref{tab:1} and Table \ref{tab:appB}.
\label{fig:Mrk841_spec}}
\end{figure}

The light curves and spectrum of the candidate event Mrk 841/2013.12.28 are shown in Fig. \ref{fig:Mrk841_lc} and Fig. \ref{fig:Mrk841_spec}. There are only two observations where the data points significantly exceed the 3$\sigma$ levels, while the hardness ratios are relatively stable in other periods. The eclipsing spectrum exhibits significant absorption in the soft X-ray, which is distinct from the spectrum of the non-eclipse phase. Because of the insufficient sampling during the eclipse-occurring period, we apply a 10-day binning time to perform time-resolved spectral analysis.

\subsection{MR 2251-178}

\begin{figure}[ht!]
\centering
\includegraphics[width=0.6\textwidth]{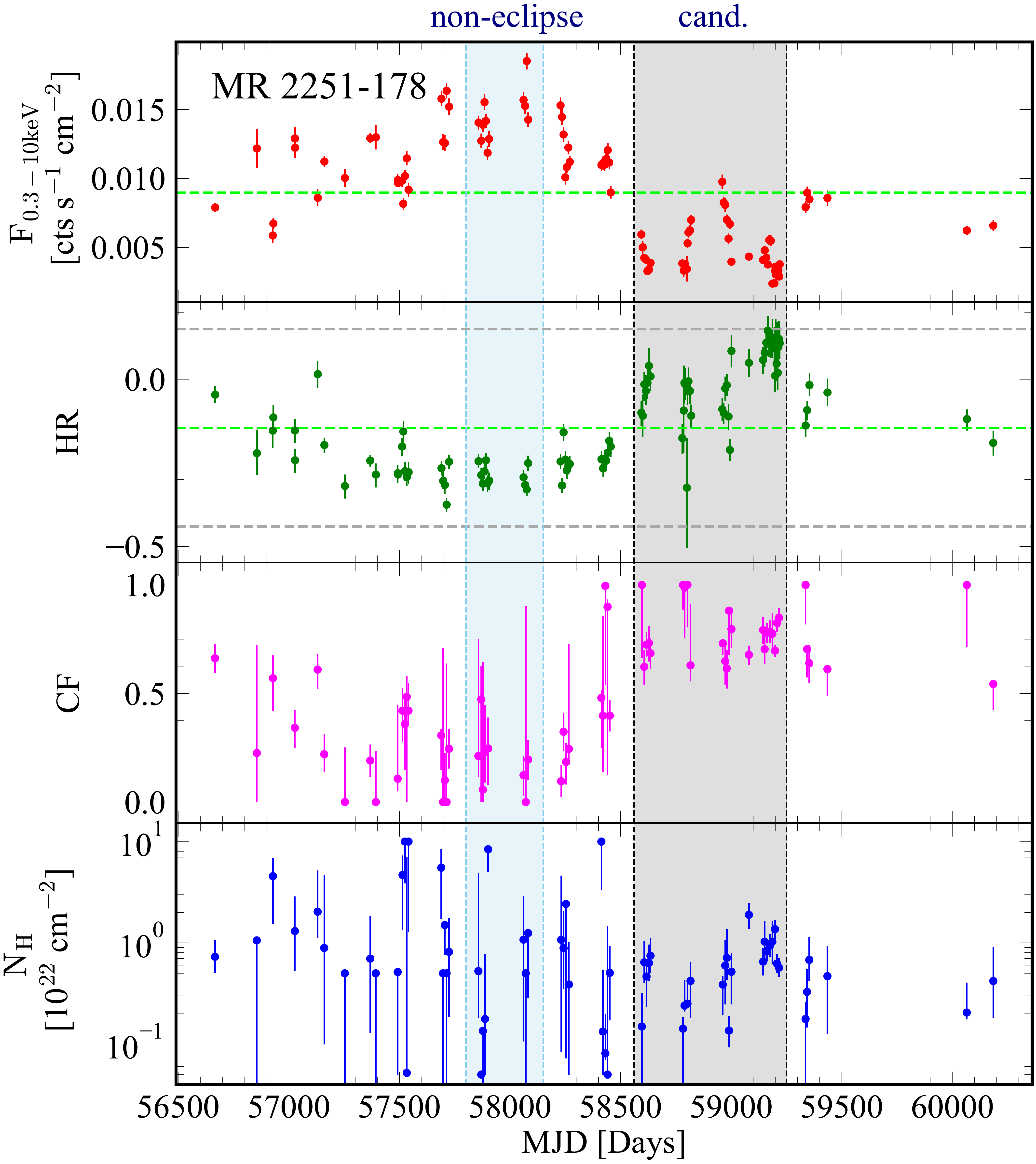}
\caption{The light curve of X-ray, hardness ratio, and time-resolved spectrum fitting parameters within 1$\sigma$ uncertainties for MR 2251-178. The time-resolved spectral fitting is performed using a 7-day bin. The highlighted regions indicate the selected periods for stacking spectra of non-eclipse phase (blue) and candidate event MR 2251-178/2020.10 (grey).
\label{fig:MR2251-178_lc}}
\end{figure}

\begin{figure}[ht!]
\centering
\includegraphics[width=0.45\textwidth]{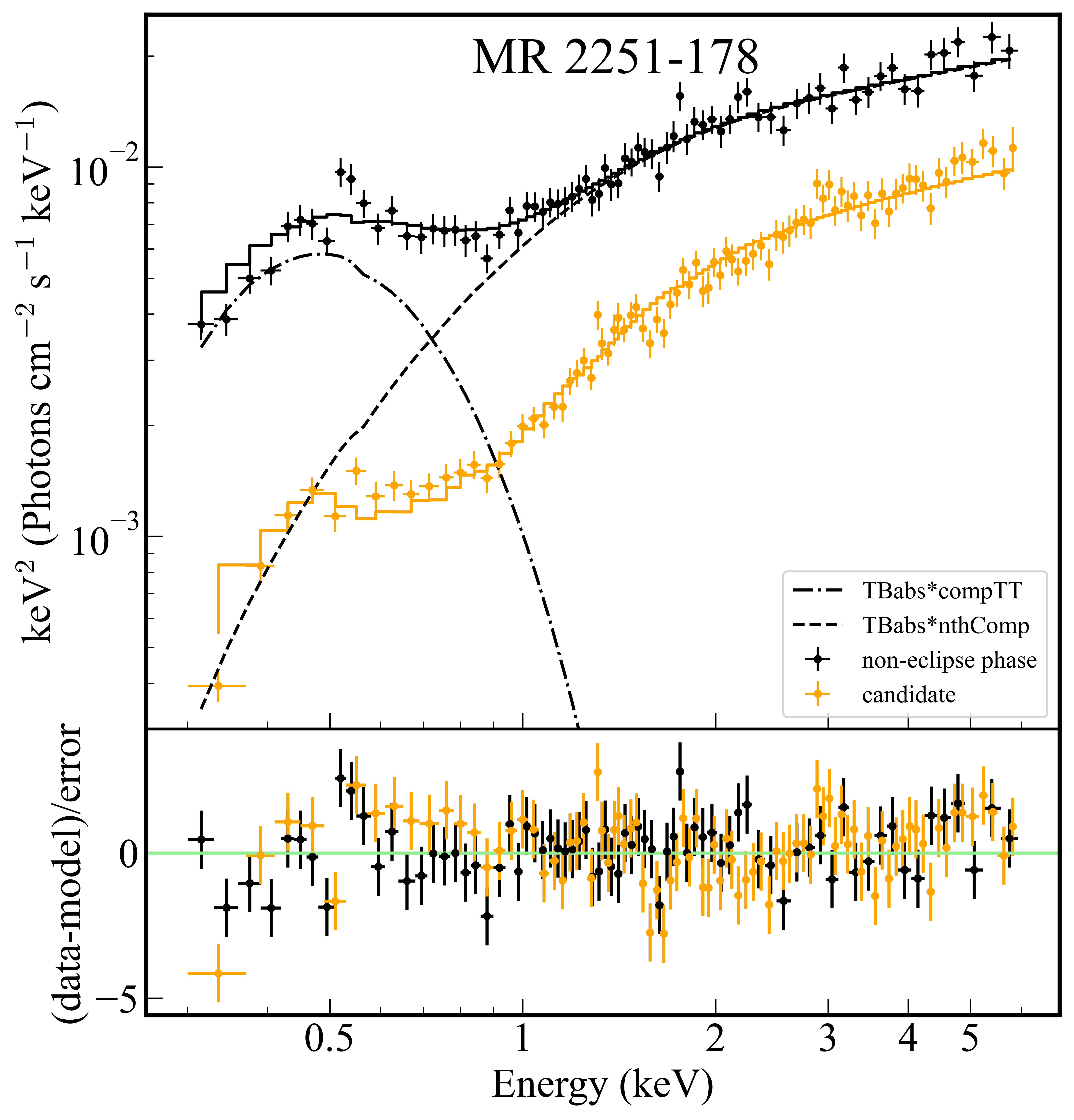}
\caption{The best fitting results of MR 2251-178 during the non-eclipse and eclipse phases, the value of parameters are listed in Table \ref{tab:1} and Table \ref{tab:appB}.
\label{fig:MR2251-178_spec}}
\end{figure}

We chose a 7-day bin to analyze the time-resolved spectra, the light curve and spectrum of candidate event MR 2251-178/2020.10 are displayed in Fig. \ref{fig:MR2251-178_lc} and \ref{fig:MR2251-178_spec}. The X-ray flux and hardness ratio show a large amplitude during the entire observation. The $N_{\rm H}$ anc CF are well-constrained during the eclipse event (highlighted in a grey shaded area), which suggests the existence of partial covering absorbers. This candidate event was also identified and further studied by \citet{2022ApJ...940...41M} using the multiple frequency observations from infrared-ray to hard X-ray.

\begin{deluxetable}{CCCCCCCCC}
\renewcommand{\arraystretch}{1.5}
\setlength{\tabcolsep}{5pt}
\tablewidth{1\textwidth}
\tablenum{B} 
\tablecaption{Best-fit parameters for the stacked spectra in the non-eclipse phases using model \texttt{TBabs*(compTT + nthComp)} with 1$\sigma$ uncertainties.
\label{tab:appB}}
\tablehead{
\colhead{Source Name} & \colhead{TBabs} & \multicolumn{5}{c}{------------------------------------   compTT   ------------------------------------} & \multicolumn{2}{c}{------------  nthComp  ------------} \\
 & \colhead{$N_{\rm H.gal}$} & \colhead{z} & \colhead{T0} & \colhead{kT} & \colhead{$\rm \tau_{p}$} & \colhead{norm} & \colhead{$\Gamma$} & \colhead{norm} \\
\colhead{(1)} & \colhead{(2)} & \colhead{(3)} & \colhead{(4)} & \colhead{(5)} & \colhead{(6)} & \colhead{(7)} & \colhead{(8)} & \colhead{(9)}
} 
\startdata
\textnormal{Mrk 817} & \text{1.18E-02} & 0.031 & 0.037$^{+0.019}_{-0.018}$ & 0.129$^{+0.012}_{-0.028}$ & 24.653$^{+13.492}_{-3.615}$ & 0.757$^{}_{-0.316}$ & 1.986$^{+0.061}_{-0.019}$ & 4.955$^{+0.188}_{-0.236}$\text{E-03}\\
\textnormal{1H 0707-495} & \text{6.55E-02} & 0.041 & 0.136$^{+0.004}_{-0.003}$ & 0.425$^{+0.106}_{-0.044}$ & 1.962$^{b}$ & 3.563$^{+0.520}_{-0.832}$\text{E-02} & 2.801$^{+1.073}_{-0.305}$ & 2.386$^{+0.269}_{-0.464}$ \text{E-04}\\
\textnormal{Mrk 509} & \text{5.04E-02} & 0.034 & 0.077$^{+0.014}_{-0.090}$ & 0.353$^{+0.0905}_{-0.102}$ & 10.322$^{+0.676}_{-1.348}$ & 0.293$^{+0.087}_{-0.0481}$ & 1.829$^{+0.065}_{-0.058}$ & 7.357$^{+2.662}_{-1.916}$ \text{E-03}\\
\textnormal{NGC 6814} & \text{15.3E-02} & 0.005 & 0.117$^{+0.080}_{-0.029}$ & 0.281$^{b}$ & 10.690$^{+3.694}_{}$ & 0.122$^{+0.023}_{-0.016}$ & 1.594$^{+0.099}_{-0.101}$ & 6.682$^{+0.627}_{-0.654}$ \text{E-03}\\
\textnormal{NGC 3227} & \text{2.12E-02} & 0.004 & 0.122$^{+0.007}_{-0.017}$ & 0.338$^{+0.023}_{-0.030}$ & 2.345$^{+4.748}_{}$ & 2.247$^{+0.272}_{-0.215}$\text{E-02} & 1.610$^{+0.065}_{-0.056}$ & 5.043$^{+0.254}_{-0.134}$ \text{E-03}\\
\textnormal{NGC 3783}$^{a}$ & \text{13.8E-02} & 0.010 & 0.094$^{+0.012}_{-0.011}$ & 0.101$^{b}$ & 2.000$^{b}$ & 0.950$^{+0.272}_{-0.271}$ & 1.573$^{+0.050}_{-0.031}$ & 1.337$^{+0.066}_{-0.047}$\text{E-02}\\
\textnormal{Mrk 841} & \text{2.43E-02} & 0.036 & 0.108$^{+0.185}_{-0.016}$ & 0.261$^{+0.067}_{-0.032}$ & 3.361$^{+20.206}_{}$ & 0.131$^{+0.046}_{-0.024}$ & 1.877$^{+0.074}_{-0.065}$ & 3.875$^{+0.272}_{-0.666}$ \text{E-03}\\
\textnormal{MR 2251-178} & \text{2.67E-02} & 0.064 & 0.116$^{+0.006}_{-0.008}$ & 0.372$^{+0.032}_{-0.033}$ & 2.162$^{+2.556}_{}$ & 0.104$^{+0.009}_{-0.017}$ & 1.679$^{+0.054}_{-0.085}$ & 6.402$^{+0.248}_{-0.306}$ \text{E-03}\\
\enddata
\tablecomments{(1)The AGNs in which eclipse events are identified in this work. (2)The equivalent hydrogen column (10$^{22}$ cm$^{-2}$) from \texttt{TBabs} model, fixed to the value of Galactic column density in the source coordinate. \\
We apply the \texttt{compTT} model to represent the warm corona and set the \texttt{approx} parameter to 1 to denote the soft photons that come from the disk. The other parameters are listed below:  (3)The redshift of each source, fixed during spectral fitting. (4)The soft photon temperature (keV). (5)The plasma temperature (keV). (6)The plasma optical depth. In some cases, only 1$\sigma$ upper errors are reported for parameters due to non-convergent fits. (7)The normalization. In Mrk 817, only 1$\sigma$ lower error is reported due to non-convergent fits.\\
We apply \texttt{nthComp} model to represent the hot corona. The electron temperature (kT$_e$) is fixed to 200 keV and the seed photon temperature is bonded to the value of plasma temperature (kT) from \texttt{compTT} model. We set the \texttt{inp$_{\rm type}$} to 0 for black body seed photons. The other parameters are listed below: (8)The asymptotic power-law photon index. (9)The normalization.\\
$^{a}$ In NGC 3783/2022.12, the non-eclipse phase also exhibits a significant absorption feature, which can be characterized by adding a higher ionized gas (\texttt{zxipcf} model) with log$\xi$ of $\sim$ 1.816, $N_{\rm H}$ of $\sim$ 2.556 $\times$ 10$^{22}$ cm$^{-2}$ and a covering factor close to 1. \\
$^{b}$ Due to the degeneracy between model components, the statistical uncertainties for these parameters could not be reliably calculated. We therefore report only the best-fit values.}
\end{deluxetable}

\begin{deluxetable}{CCCCCC}
\renewcommand{\arraystretch}{1.5}
\setlength{\tabcolsep}{5pt}
\tablewidth{1\textwidth}
\tablenum{C}
\tablecaption{The parameter ranges of $N_{\text H}$, CF, and constant factor during eclipse periods, derived from time-resolved spectral fitting using model \texttt{TBabs*zxipcf*constant*(compTT + nthComp)}. 
\label{tab:appC}}
\tablehead{
\colhead{Event} & \multicolumn{4}{c}{---------------------------  zxipcf  --------------------------} & \colhead{constant} \\
 & \colhead{$N_{\text H}$} & \colhead{\text{log}\,$\xi$} & \colhead{CvrFract} & \colhead{z} & \colhead{factor} \\
\colhead{(1)} & \colhead{(2)} & \colhead{(3)} & \colhead{(4)} & \colhead{(5)} & \colhead{(6)}
} 
\startdata
\textnormal{Mrk 817/2018.3}$^a$ & 1.02 $-$ 23.08 & 0.903 & 0.17 $-$ 0.84 & 0.031 & 0.28 $-$ 1.50 \\
\textnormal{Mrk 817/2018.5}$^b$ & 1.08 $-$ 19.58 & 0.708 & 0.1 $-$ 0.85 & 0.031 & 0.64 $-$ 1.44 \\
\textnormal{1H 0707/2010.11}$^b$ & 1.96 $-$ 64.50 & 1.936 & 0.38 $-$ 1.00 & 0.041 & 0.10 $-$ 2.50 \\
\textnormal{Mrk 509/2017.5}$^b$ & 0.14 $-$ 0.42 & $-$0.547 & 0.24 $-$ 0.76 & 0.034 & 0.65 $-$ 0.89 \\
\textnormal{NGC 6814/2016.4}$^a$ & 0.39 $-$ 5.40 & 2.102 & 0.29 $-$ 1.00 & 0.005 & 0.51 $-$ 1.26 \\
\textnormal{NGC 3227/2008.10}$^a$ & 5.19 $-$ 19.46 & 1.966 & 0.64 $-$ 0.91 & 0.004 & 0.57 $-$ 1.10 \\
\textnormal{NGC 3227/2019.10}$^a$ & 0.96 $-$ 5.55 & 0.754 & 0.44 $-$ 0.83 & 0.004 & 0.39 $-$ 0.72 \\
\textnormal{NGC 3783/2022.12}$^a$ & 1.34 $-$ 8.52 & $-$0.89 $-$ 1.60 & 0.75 $-$ 1.00 & 0.010 & 0.71 $-$ 1.15 \\
\textnormal{MR 2251/2020.10}$^a$ & 0.42 $-$ 1.36 & $-$0.572 & 0.62 $-$ 0.85 & 0.064 & 0.34 $-$ 0.76 \\
\enddata
\tablecomments{For the candidates NGC 6814/2016.3.30 and Mrk 841/2013.12.28, the time-resolved spectral analysis cannot effectively capture their eclipse processes, because {\it Swift} had fewer than four observations during their peak eclipse states. As a result, we do not list the parameter ranges of these events derived from the \texttt{zxipcf} model; see Table \ref{tab:1} for their average values. Intrinsic parameters are fixed to average values due to low signal-to-noise spectra, and values of log$\xi$ are also fixed to the average value, except for the event NGC 3783/2022.12. The non-eclipse phase of this source also has a significant absorption feature. It can be described by a high-ionized gas, so the range of log$\xi$ values for candidate NGC 3783/2022.12 is listed in the table. \\
(1)The identified event and their estimated starting/peaking time (UTC): $^a$The candidate eclipse event; $^b$The high-confidence eclipse event. \\
The components in \texttt{zxipcf} model: (2)The equivalent hydrogen column (10$^{22}$ cm$^{-2}$). (3)The ionization parameter (\text{erg s}$^{-1}$ \text{cm}), we fix to the best-fit value of stacked spectra in eclipse phases, except NGC 3783. (4)The covering fraction. (5)The redshift, we fix to the redshift of each source. \\
(6)The normalization factor from the \texttt{constant} model. \\
}
\end{deluxetable}




\end{document}